\documentclass[titlepage,11 pt]{article}

\usepackage[centertags]{amsmath}
\usepackage{graphicx}
\usepackage[english]{babel}
\usepackage{natbib}
\usepackage{url} 
\usepackage{pifont}
\usepackage{hyperref}
\usepackage{amsmath}
\usepackage{amsfonts}

\usepackage{multicol}
\usepackage{bbm}
\usepackage{url} 

\usepackage[affil-it]{authblk}

\usepackage{hyperref}

\usepackage{setspace}
\usepackage{lipsum}
\usepackage[margin=3cm]{geometry}

\newtheorem{thm}{Theorem}[section]
\newtheorem{coroll}{Corollary}[section]
\newtheorem{Lemma}{Lemma}[section]

\newtheorem{Def}{Definition}[section]
\newtheorem{Ass}{Assumption}[section]

\title{\textbf{A large covariance matrix estimator\\ under intermediate spikiness regimes}}

\author{\textbf{Matteo Farn\'{e}}  \thanks{Electronic address: \texttt{matteo.farne2@unibo.it}; Corresponding author}}
\affil{Department of Statistical Sciences,\\ University of
Bologna, Italy}

\author{\textbf{Angela Montanari}  
}
\affil{Department of Statistical Sciences,\\ University of
Bologna, Italy}

\begin{document}

\maketitle

\begin{abstract}
The present paper concerns large covariance matrix estimation
via composite minimization under the assumption of low rank plus sparse structure.
In this approach, the low rank plus sparse decomposition of the covariance matrix is recovered
by least squares minimization under nuclear norm plus $l_1$ norm penalization. This paper proposes a new estimator of that family based on an additional least-squares re-optimization step aimed at
un-shrinking the eigenvalues of the low rank component estimated at the first step.
We prove that such un-shrinkage causes the final estimate to approach the target as closely as possible in Frobenius norm
while recovering exactly the underlying low rank and sparsity pattern.
Consistency is guaranteed when $n$ is at least $O(p^{\frac{3}{2}\delta})$,
provided that the maximum number of non-zeros per row in the sparse component is $O(p^{\delta})$ with $\delta \leq \frac{1}{2}$.
Consistent recovery is ensured if the latent eigenvalues scale to $p^{\alpha}$, $\alpha \in[0,1]$, while rank consistency is ensured if $\delta \leq \alpha$.
The resulting estimator is called UNALCE (UNshrunk ALgebraic Covariance Estimator) and is shown to outperform state of the art estimators,
especially for what concerns fitting properties and sparsity pattern detection.
The effectiveness of UNALCE is highlighted on a real example regarding ECB banking supervisory data.
\end{abstract}

%
%
%
%
%
%
%
%


\section{Introduction}

Estimation of population covariance matrices from samples
of multivariate data is of interest in many high-dimensional inference problems -
principal components analysis,
classification by discriminant analysis, inferring a graphical model structure, and
others. Depending on the different goal the interest is sometimes in inferring the eigenstructure of the covariance matrix (as in PCA) and sometimes in estimating its inverse (as in discriminant analysis or in graphical models).
Examples of application areas where these problems arise include gene arrays,
fMRI, text retrieval, image classification, spectroscopy, climate studies, finance and macro-economic analysis.

The theory of multivariate analysis for normal variables has been well worked out
(see, for example, \cite{anderson1984multivariate}). However, it became soon apparent that exact expressions
were cumbersome, and that multivariate data were rarely Gaussian. The remedy
was asymptotic theory for large samples and fixed, relatively small, dimensions.
However, in recent years, datasets that do not fit into this framework have become very
common, since nowadays the data can be very high-dimensional and sample sizes can be very small relative to dimension.

The most traditional covariance estimator, the sample covariance matrix, is known to be dramatically ill-conditioned in a large dimensional context, where the process dimension $p$ is larger than or close to the sample size $n$, even when the population covariance matrix is well-conditioned.
Two key properties of the matrix estimation process assume a particular relevance in large dimensions:
well conditioning (i.e. numerical stability) and 
identifiability.
Both properties are crucial for the theoretical recovery and the practical use of the estimate. A bad conditioned estimate suffers from collinearity and causes its inverse, the precision matrix, to  dramatically amplify any error in the data. A large dimension may cause the impossibility to identify the unknown covariance structure thus hampering the interpretation of the results.

Regularization approaches to large covariance matrices estimation have therefore started to be presented in the literature,
both from theoretical and practical points of view (see \cite{fan2016overview} for an exhaustive overview).
Eigenvalue regularization approaches include linear \citep{ledoit2004well} and nonlinear shrinkage (\cite{ledoit2015spectrum}, \cite{lam2016nonparametric}).
Sparsity-based approaches include penalized likelihood maximization \citep{friedman2008sparse},
tapering (\cite{furrer2007estimation}, \cite{cai2010}), 
banding \citep{bickel2008regularized}
and thresholding (\cite{bickel2008covariance}, \cite{rothman2009generalized}, \cite{cai2011adaptive}).
A consistent bandwidth selection method for all these approaches is described in \cite{qiu2015bandwidth}.

A different approach is based on the assumption of a low rank plus sparse structure for the covariance matrix:
\begin{equation}
{\Sigma}^{*}={L}^{*}+{S}^{*}\label{model:base},
\end{equation}
where 
${L}^{*}$ is low rank with rank $r<p$,
${S}^{*}$ is positive definite and sparse with at most $s$ nonzero off-diagonal elements, and ${\Sigma}^{*}$ is a positive definite matrix.
The generic covariance estimator $\hat{{\Sigma}}$ can be written as
\begin{equation}
\hat{{\Sigma}}={L}^{*}+{S}^{*}+ {W}={\Sigma}^{*}+{W}\label{model:noise},
\end{equation}
where ${W}$ is an error term. 
The error matrix ${W}$ may be deterministic or stochastic, as explained in \cite{agarwal2012noisy}.
If the data are Gaussian and $\hat{{\Sigma}}$ is the unbiased sample covariance matrix ${\Sigma}_n$,
${W}$ is distributed as a re-centered Wishart.

In \cite{fan2013large}, a large covariance matrix estimator, called POET (Principal Orthogonal complEment Thresholding), is derived under this assumption.
POET combines Principal Component Analysis for the recovery of the low rank component and a thresholding algorithm for the recovery of the sparse component.
The underlying model assumptions prescribe an approximate factor model with spiked eigenvalues (i.e. growing with $p$) for the data, thus allowing to reasonably use the truncated PCA of the sample covariance matrix. Furthermore, at the same time, sparsity in the sense of \cite{bickel2008covariance} is imposed to the residual matrix.
The latent rank $r$ is chosen by the information criteria of \cite{bai2002determining}.

Indeed, rank selection represents a relevant issue: if $p$ is large, setting a large rank would cause the estimate $\hat{{\Sigma}}$ to be non-positive definite, while setting a small rank would cause a too relevant variance loss. In the discussion of \cite{fan2013large}, Yu and Samworth point out that the probability to underestimate the latent rank does not asymptotically vanish if the eigenvalues are not really spiked at rate $O(p)$.
In addition, we note that POET systematically overestimates the proportion of variance explained by the factors (given the true rank)
because the eigenvalues of ${\Sigma}_n$ are more spiky than the true ones (as showed in \cite{ledoit2004well}).

POET asymptotic consistency holds given that a number of assumptions is satisfied. The key assumption is the pervasiveness of latent factors, which causes the PCA of ${\Sigma}_n$ to asymptotically identify the eigenvalues and the eigenvectors of ${\Sigma}^{*}$ as $p$ diverges. The results of \cite{fan2013large} provide the convergence rates of the relative norm of $\hat{{\Sigma}}_{POET}-{\Sigma}^{*}$ (defined as $||\hat{{\Sigma}}_{POET}-{\Sigma}^{*}||_{\Sigma}=p^{-1/2}||{\Sigma}^{*-\frac{1}{2}}\hat{{\Sigma}}_{POET} {\Sigma}^{*-\frac{1}{2}}~-~{I}_p||_{Fro}$), the maximum norm of $\hat{{\Sigma}}_{POET}-{\Sigma}^{*}$ and the spectral norm of $\hat{{S}}_{POET}-{S}^{*}$. Under stricter conditions, $\hat{{S}}_{POET}$ and $\hat{{\Sigma}}_{POET}$ are proved to be non-singular with probability approaching $1$.

At the same time, a number of non-asymptotic methods has been presented.
In \cite{chandrasekaran2011rank} the exact recovery of the covariance matrix in the noiseless context is first proved. The result is achieved minimizing a specific convex non-smooth objective, which is the sum of the nuclear norm of the low rank component and the $l_1$ norm of the sparse component. In \cite{chandrasekaran2012}, which is  an  extension of \cite{chandrasekaran2011rank}, the exact recovery of the inverse covariance matrix by the same numerical problem in the noisy graphical model setting is provided.
The authors prove that, in the worst case, the number of necessary samples in order to ensure consistency is $n=O\left({p^3}/{r^2}\right)$,
even if 
the required condition for the
positive definiteness of the estimate is $p\leq 2n$.

An approximate solution to the recovery and identifiability
of the covariance matrix in the noisy context is described in \cite{agarwal2012noisy}.
Even there, the condition $p \leq n$ is unavoidable, for standard results on large deviations and non-asymptotic random matrix theory.
An exact solution to the same problem, based on the results in \cite{chandrasekaran2012}, is then shown in \cite{luo2011recovering}.
The resulting estimator is called LOREC (LOw Rank and sparsE Covariance estimator) and is proved to be both
algebraically and parametrically consistent in the sense of \cite{chandrasekaran2012}.

In \cite{chandrasekaran2012} algebraic consistency is defined as follows
\begin{Def}
A pair of symmetric matrices $({S},{L})$ with ${S}, {L} \in R^{p\times p}$ is an algebraically consistent estimate of the low rank plus sparse model (\ref{model:noise}) for the covariance matrix ${\Sigma}^{*}$ if the following conditions hold: 
\begin{enumerate}
\item The sign pattern of ${S}$ is the same of ${S}^{*}$:
$sign({S}_{ij})=sign(({S}^{*})_{i,j})$, $\forall i,j$.
Here we assume that $sign (0)=0$.
\item The rank of ${L}$ is the same as the rank of ${L}^{*}$.
\item Matrices ${L}+{S}$, ${S}$ and ${L}$ are such that ${L}+{S}$ and ${S}$ are positive definite
and ${L}$ is positive semidefinite.
\end{enumerate}
\end{Def}
Parametric consistency holds if the estimates of $({S},{L})$ are close to $({S}^{*},{L}^{*})$ in some norm
with high probability. In \cite{chandrasekaran2012} such norm is\\ $g_\gamma=\max\left(\frac{||\hat{{S}}-{S}^{*}||_{\infty}}{\gamma},||\hat{{L}}-{L}^{*}||_2\right).$

LOREC shows several advantages respect to POET.
The most important is that the estimates are both algebraically and parametrically consistent,
while POET provides only parametric consistency.
In spite of that, LOREC suffers from some drawbacks, especially concerning fitting properties.
What is more, the strict condition $p\leq n$ is required, while POET allows for $p\log(p)\gg n$.

For these reasons, we propose a new estimator, UNALCE (UNshrunk ALgebraic Covariance Estimator), based on the unshrinkage
of the estimated eigenvalues of the low rank component, which allows to improve the fitting properties of LOREC systematically.
We assume that the non-zero eigenvalues of ${L}^{*}$ and ${\Sigma}^{*}$ are proportional to $p^{\alpha}$, $\alpha \in [0,1]$ (the so called generalized spikiness context).
%
Under the assumption that the maximum number of non-zeros per row in ${S}^*$, called ''maximum degree'', is $O(p^{\delta})$ (with $\delta \leq \frac{1}{2}$),
we prove that our estimator possesses a non-asymptotic error bound admitting that $n$ is as small as $O(p^{3 \delta})$.
We derive absolute bounds depending on $\alpha$ for the low rank, the sparse component, and the overall estimate,
as well as the conditions for rank consistency, positive definiteness and invertibility.
In this way we provide a unique framework for covariance estimation via composite minimization under the low rank plus sparse assumption.

The remainder of the paper is organized as follows.
In Section \ref{unshr} we first define ALCE (ALgebraic Covariance Estimator) with the necessary assumptions for algebraic and parametric consistency, and then define UNALCE, proving that the unshrinkage of thresholded eigenvalues of the low rank component is the key to improve fitting properties as much as possible given a finite sample, preserving algebraic consistency.
In Section \ref{crit} we propose a new model selection criterion specifically tailored to our model setting.
In Section \ref{real} we provide a real Euro Area banking data example which clarifies the effectiveness of our approach. Finally, in Section \ref{concl} we draw the conclusions and discuss the most relevant findings.



\section{Numerical estimation and spiked eigenvalues: the ALCE approach}\label{unshr}


\subsection{The model}\label{data}

First of all, we recall the definitions of the matrix norms used throughout the paper.
Let us define a $p\times p$ symmetric positive-definite matrix ${M}$.
We denote by $\lambda_i({M})$, $i=1,\ldots,p$, the eigenvalues of ${M}$ in descending order.
Then we recall the following norms definitions:
\begin{enumerate}
\item element-wise:
\begin{enumerate}
\item $L_0$ norm: $||{M}||_0=\sum_{i=1}^p \sum_{j=1}^p \mathbbm{1}(m_{ij}\ne 0)$, which is the total number of non-zeros.
\item $L_1$ norm: $||{M}||_1=\sum_{i=1}^p \sum_{j=1}^p |m_{ij}|$;
\item Frobenius norm: $||{M}||_{Fro}=\sum_{i=1}^p \sum_{j=1}^p m_{ij}^2$;
\item maximum norm: $||{M}||_{\infty}=\max_{i\leq p, j \leq p} |m_{ij}|$;
\end{enumerate}
\item induced by vector:
\begin{enumerate}
\item $||{M}||_{0,v}=\max_{i \leq p} \sum_{j \leq p} \mathbbm{1}(m_{ij} \ne 0)$, which is the maximum number of non-zeros per column,
defined as the maximum ''degree'' of ${M}$; 
\item $||{M}||_{1,v}=\max_{i \leq p} \sum_{j \leq p} |m_{ij}|$;
\item spectral norm: $||{M}||_{2}=\lambda_1({M})$;
\end{enumerate}
\item Schatten:
\begin{enumerate}
\item nuclear norm of ${M}$, here defined as the sum of the eigenvalues of ${M}$:
$||{M}||_{*}=\sum_{i=1}^p \lambda_i({M})$.
\end{enumerate}
\end{enumerate}

Let us suppose the population covariance matrix of our data is the sum of a low rank and a sparse component.
A $p$-dimensional random vector ${x}$ is said to have a \textbf{low rank plus sparse structure} if its covariance matrix ${\Sigma}^{*}$ satisfies the following relationship:

\begin{equation}
{\Sigma}^{*}={L}^{*}+{S}^{*},\label{mod:good}
\end{equation} where:
\begin{enumerate}
\item ${L}^{*}$ is a positive semidefinite symmetric $p \times p$
matrix with at most rank $r \ll p$; \item ${S}^{*}$ is a positive definite
$p \times p$ sparse matrix with at most $s \ll p(p-1)/2$ nonzero off-diagonal
elements and maximum degree $s'$.
\end{enumerate}

According to the spectral theorem, we can write ${L}^{*}={U_LDU_L'}={BB'}$, where ${B}={U_L D^{1/2}}$,
${U_L}$ is a $p \times r$ semi-orthogonal matrix, ${D}$ is a $r \times r$ diagonal matrix,
with $d_{jj}>0$, $\forall j=1, \ldots, r$.
Let us suppose that the $p\times 1$ random vector ${x}$ is generated according to the following model:
\begin{equation}
{x}={B} {f}+{\epsilon},
\end{equation}
where ${f}$ is a $r \times 1$ random vector with
$E({f})={0_r}$, $V({f})={I_r}$ and 
${\epsilon}$ is $p \times 1$ random vector with $E({\epsilon})={0_p}$,$V({\epsilon})={S}^{*}$.
The random vector ${x}$ is thus assumed to be zero mean, without loss of generality. Given a sample ${x}_i$, $i=1,\ldots,n$,
${\Sigma}_n=\frac{1}{n-1}\sum_{i=1}^n {x}_i {x}_i'$ is the $p \times p$ sample covariance matrix.

It is easy to observe that ${x}$ follows a low rank plus sparse structure:
\begin{eqnarray}
E({x}{x}')=E\left\{({B} {f}+{\epsilon})({B} {f}+{\epsilon})'\right\}=\nonumber\\
=E({B}' {f}' {f} {B}) + E({B} {f} {\epsilon}') + E({\epsilon} {B}'{f}')+ E({\epsilon} {\epsilon}')=\\ \nonumber
={BB}'+{S}^{*}={L}^{*}+{S}^{*}={\Sigma}^{*}
\end{eqnarray}
under the usual assumption ${f} \perp {\epsilon}$, i.e. $cov({f},{\epsilon})=E({f} {\epsilon}')=E({\epsilon} {f}')={0}_{r \times p}$
($r\times p$ null matrix).
Assuming $p\leq n$, it is also useful to recall that for $n \rightarrow \infty$
\begin{eqnarray}
E({\Sigma}_n)=E\left\{\frac{1}{n-1}{x}{x}'\right\}=E\left\{\frac{1}{n-1}({B} {f}+{\epsilon})({B} {f}+{\epsilon})'\right\}=\nonumber\\
=E\left\{\frac{1}{n-1}\left({B}' {f}' {f} {B} +{B} {f} {\epsilon}' + {\epsilon} {B}'{f}'+ {\epsilon} {\epsilon}'\right)\right\}=\\ \nonumber
={BB}'+{S}^{*}={L}^{*}+{S}^{*}={\Sigma}^{*}
\end{eqnarray}
If we assume a normal distribution for ${f}$ and ${\epsilon}$, the above equality is true  for any fixed $n$ and the matrix
${W}:={\Sigma}_n~-~({L}^{*}+{S}^{*})$ is distributed as a re-centered Wishart noise.
In any case, the normality assumption is not essential for our setting.

\subsection{Nuclear norm plus $l_1$ norm heuristics}\label{sec:LOREC}

Under model (\ref{model:noise}), the need rises to develop a method able at the same time to consistently estimate the covariance matrix ${\Sigma}^{*}$ as well as to catch the sparsity pattern of ${S}^{*}$ and the spikiness pattern of the eigenvalues of ${L}^{*}$ simultaneously.
Such estimation problem is stated as
\begin{equation}
\min_{{L},{S}} \frac{1}{2}||({L}+{S})-{\Sigma}_n||_{Fro}^2 +  \psi
rank({L}) + \rho ||{S}||_{0,off},\label{prob:orig1}
\end{equation}
where $||{S}||_{0,off}=\sum_{i=1}^{p-1} \sum_{j=i+1}^{p} \mathbbm{1}(s_{ij}^{*}\ne 0)$ (because the diagonal of ${S}$ is preserved as in \cite{fan2013large}).
This is a combinatorial problem, which is known to be NP-hard, since both $rank({L})$ and $||{S}||_{0,off}$ are not convex.


The tightest convex relaxation of problem (\ref{prob:orig1}), as shown in \cite{fazel2002matrix}, is
\begin{equation}
\min_{{L},{S}} \frac{1}{2}||({L}+{S})-{\Sigma}_n||_{Fro}^2 +  \psi
||{L}||_{*} + \rho ||{S}||_{1,off},\label{func:ob1}%
\end{equation}
where $\psi$ and $\rho$ are non-negative \textbf{threshold} parameters, and $||{S}||_{off,1}=\sum_{i=1}^{p-1} \sum_{j=i+1}^p |s_{ij}^{*}|$.
The use of nuclear norm for covariance matrix estimation was introduced in \cite{fazel2001rank}. 
The feasible set of (\ref{func:ob1}) is the set of all $p\times p$ positive definite matrices ${S}$ and all $p\times p$ positive semi-definite matrices ${L}$.


From a statistical point of view, (\ref{func:ob1}) is a penalized least squares heuristics, composed by a smooth least squares term ($\frac{1}{2}||({L}+{S})-{\Sigma}_n||_{Fro}^2$) and a non-smooth composite penalty ($\psi
||{L}||_{*} + \rho ||{S}||_{1}$).
The choice of (\ref{func:ob1}) allows to lower the condition number of the estimates
and the parameter space dimensionality simultaneously.
The optimization of (\ref{func:ob1}) requires the theory of non-smooth convex optimization provided by \cite{rockafellar2015convex} and \cite{clarke1990optimization} (the solution algorithm is reported in the Supplement).

In principles, different losses could be used, like Stein's one \citep{dey1985estimation}.
However, the classical Frobenius loss does not require normality and is computationally appealing. The study of different fitting terms, including the ones performing eigenvalue regularization, is left to future research.

From an algebraic point of view, (\ref{func:ob1}) is an algebraic matrix variety recovery problem.
In the noisy covariance matrix setting described in equation (\ref{mod:good}), matrices ${L}^{*}$ and ${S}^{*}$
are assumed to come from the following sets of matrices:
\begin{eqnarray}
\mathcal{B}(r) & = & \{{L} \in R^{p \times p} \mid {L}={UDU'}, {U} \in R^{p
\times r} \mbox{semi-orthogonal}, {D} \in R^{r \times r} \mbox{diagonal}\}\label{var:L}\\
\mathcal{A}(s) & = & \{{S}\in R^{p\times p} \mid |support({S})| \leq
s\}.\label{var:S}
\end{eqnarray}
$\mathcal{B}(r)$ is the variety of matrices with \textbf{at most}
rank $r$.
$\mathcal{A}(s)$ is the variety of (element-wise) sparse matrices with 
\textbf{at most} $s$ nonzero elements, where $support({S})$ is the orthogonal complement of $ker({S})$. 
%



In \cite{chandrasekaran2011rank} the notion of rank-sparsity incoherence is developed, which is defined as the uncertainty principle between the sparsity pattern of a matrix and its row/column space. Denoting by $T({L})$ and $\Omega({S})$ the tangent spaces to $\mathcal{B}(r)$ and $\mathcal{A}(s)$ respectively,
the following rank-sparsity incoherence measures between $\Omega({S}^{*})$ and $T({L}^{*})$ are defined:
\begin{eqnarray}
\xi(T({L}^{*})) & = & \max_{{N} \in T({L}^{*}), ||{N}||_2 \leq 1} {||{N}||_\infty}, \label{xi}\\
\mu(\Omega({S}^{*})) & = &\max_{{N} \in \Omega({S}^{*}),||{N}||_\infty \leq 1}\
{||{N}||_2}\label{mu}.
\end{eqnarray}
In order to identify $T({L}^{*})$ and $\Omega({S}^{*})$, we need quantities $\xi(T({L}^{*}))$
and $\mu(\Omega({S}^{*}))$ to be as small as possible, because the smaller they are, the better is the decomposition.
The product $\mu(\Omega({S}^{*}))\xi(T({L}^{*}))$ is the rank-sparsity incoherence measure
and bounding it controls both for identification and recovery.

The described approach was first used for deriving LOREC estimator in \cite{luo2011recovering}.
Therein, the reference matrix class imposed to ${\Sigma}^{*}$ is
\begin{equation}
{\Sigma}^{*}( \epsilon_0)=\{{\Sigma}^* \in
R^{p \times p}: 0< \epsilon_0\leq \lambda_i({\Sigma}^{*})\leq  \epsilon_0^{-1}, \, \forall i=1,\ldots,p\}\label{def:U_eps}
\end{equation}
which is the class of positive definite matrices having uniformly bounded eigenvalues.
In the context so far described, Luo proves that ${L}$ and ${S}$ can be identified and recovered with bounded error, and the rank of ${L}$ as well as the sparsity pattern of ${S}$ are exactly recovered. 

The key model-based results for deriving LOREC consistency bounds are
a lemma by \cite{bickel2008covariance} for the sample loss in infinity (element-wise) norm:\\
\begin{equation}||{\Sigma}_n-{\Sigma}^{*}||_{\infty} = O\left(\sqrt{\frac{\log{p}}{n}}\right),\label{BLrate}\end{equation}
and
a lemma by  \cite{davidson2001local} for the sample loss in spectral norm:\\
\begin{equation}||{\Sigma}_n-{\Sigma}^{*}||_2 = O\left(\sqrt{\frac{p}{n}}\right).\label{DSrate}\end{equation}
We stress that (\ref{DSrate}) strictly requires the assumption $p \leq n$. 

From a theoretical point of view, LOREC approach presents some deficiencies and incongruities.
Differently from POET approach, where the sparsity assumption is imposed to the sparse component ${S}^{*}$, LOREC approach imposes it directly to the covariance matrix ${\Sigma}^{*}$. As a consequence, the assumption ${\Sigma}^{*} \in {\Sigma}^{*}(\epsilon_0)$ (see (\ref{def:U_eps})) is necessary and causes, jointly with the identifiability assumptions, uncertainty on the underlying structure of ${\Sigma}^{*}$.

In fact, assuming uniformly bounded eigenvalues may conflict with the main necessary identifiability condition: the transversality between $\Omega$ and $T$. Since the eigenvalue structures of ${\Sigma}^{*}$ and ${S}^{*}$ are somehow linked, requiring class (\ref{def:U_eps}) for ${\Sigma}^{*}$ may cause ${S}^{*}$ to be not enough sparse, and simultaneously the row/column space of ${L}^{*}$ to have high values of incoherence, because we have no spiked eigenvalues. This may result in possible non-identifiability issues.

\subsection{ALCE estimator}\label{sec:ALCE}
Let us suppose that the eigenvalues of ${\Sigma}^{*}$ are intermediately spiked with respect to $p$.
This equals to assume the generalized spikiness of latent eigenvalues in the sense of Yu and Samworth (\cite{fan2013large}, p. 656):
\begin{Ass}\label{spikyPOETnew}
All the eigenvalues of the $r\times r$ matrix $p^{-\alpha}B'B$ are bounded away from $0$ for all $p$ and $\alpha \in [0,1]$.
\end{Ass}
If $p$ is finite, Assumption \ref{spikyPOETnew} is equivalent to state that 
\begin{eqnarray}
\lambda_{1, \ldots, r}({\Sigma}^{*}) & \geq & \delta_{\alpha} p^{\alpha},\nonumber\\
\lambda_{r+1, \ldots, p}({\Sigma}^{*}) & \leq & \delta_{\alpha} p^{\alpha},\nonumber
\end{eqnarray}for some $\delta_{\alpha} > 0$.
Hence, we aim to study the properties of the covariance estimates obtained by heuristics (\ref{func:ob1})
under the generalized spikiness assumption in a non-asymptotic context.
%


In order to do that, we need to study the behaviour of the model-based quantity $P(||{\Sigma}_n~-~{\Sigma}^{*}||)$,  
which is the only probabilistic component.
We bound $P(||{\Sigma}_n-{\Sigma}^{*}||_{\infty})$ exploiting the property $||.||_{\infty}\leq ||.||_2$.
Therefore, our aim is to show that 
\begin{equation}
P\left(||{\Sigma}_n-{\Sigma}^{*}||> C_1\frac{p^{\alpha}}{\sqrt{n}}\right)\leq  1- C_2 \exp{(-C_3 p^{2\alpha})}\label{plusmore},
\end{equation}
which is verified if it holds
\begin{equation}\label{samplePOET}
||{\Sigma}_{n}-{\Sigma}^{*}|| \leq C \frac{p^{\alpha}}{\sqrt{n}}
\end{equation}
with very high probability ($C_1$, $C_2$, $C_3$ and $C$ are positive constants).
Exploiting the consistency norm 
of \cite{chandrasekaran2012}, which is \begin{equation}g_\gamma=\max\left(\frac{||\hat{{S}}-{S}^{*}||_{\infty}}{\gamma},||\hat{{L}}-{L}^{*}||_2\right),\label{ggamma2}\end{equation}
it follows from (\ref{samplePOET}) that 
\begin{equation} g_\gamma(\hat{{S}}-{S}^{*},\hat{{L}}-{L}^{*})\leq C \frac{1}{\xi(T)}{\frac{p^{\alpha}}{\sqrt{n}}}\label{tiprego}\end{equation}
with very high probability (see \cite{luo2011recovering} for technical details).

In order to reach this goal,
we need to impose that the following assumptions hold in our finite sample context.

\begin{Ass}\label{alg}
There exist $k_L,k_S>0$, $\delta\leq \frac{1}{2}$, such that $\xi(T({L}))=\sqrt{\frac{r}{k_L^2 p^{2\delta}}}$, $\mu(\Omega({S}))=k_S p^{\delta}$,
$\frac{k_S}{k_L}\leq \frac{1}{54}$ with $\delta \leq \alpha$.
\end{Ass}

\begin{Ass}\label{tails}
There are $r_1,r_2>0$ and $b_1,b_2>0$ such that, for any $s>0$, $i\leq n$, $j \leq r$, $j'\leq p$:
\begin{eqnarray}
P(|f_{ij}|>s) & \leq & \exp({-b_1/s}), \nonumber \\
P(|\epsilon_{ij'}|>s) & \leq & \exp({-b_2/s}). \nonumber
\end{eqnarray}
\end{Ass}

\begin{Ass}\label{tails2}
There are constants $c_1,c_2,c_3,\delta_2>0$ such that $\lambda(S^*)_{min} > c_1$,\\ $\min_{i,i' \leq p}var(\epsilon_{ij}\epsilon_{i'j})>c_2$
for any $j\leq n$, $i \leq r$, $i'\leq p$, $s_{ii} \leq c_3 p^{\delta}$,\\
and $s'=\max \sum_{j \leq p} \mathbbm{1}(s^*_{ij}=0)\leq \delta_2 p^{\delta}$, $\delta_2\geq k_S$.
\end{Ass}

\begin{Ass}\label{pr}
There exist $\delta_3,\delta_4>0$ such that $r\leq \delta_3 \log{p^{3\delta}}$ and $n\geq \delta_4 p^{\frac{3}{2}\delta}$.
\end{Ass}

\begin{Ass}\label{sc}
$\alpha\leq 3\delta$ and $\frac{1}{k_L\delta_4}<\delta_2$.
\end{Ass}

Assumption \ref{alg} is needed to ensure algebraic consistency.
In fact, an identifiability condition for problem (\ref{func:ob1}), as shown in Theorem \ref{thmMinetop}, is $\xi(T({L}^*))\mu(\Omega({S}^*))\leq \frac{1}{54}$. 
According to \cite{chandrasekaran2011rank}, it holds $\sqrt{\frac{r}{p}} \leq \xi(T({L}^*)) \leq 1$ and
$\min \sum_{j \leq p} \mathbbm{1}(s^*_{ij} \ne 0) \leq \mu(\Omega({S}^*)) \leq \max \sum_{j \leq p} \mathbbm{1}(s^*_{ij} \ne 0)$.
It descends that $\xi(T({L}^*))=1$ with $\delta=0$ in the worst case scenario
and $\xi(T({L}^*))=\sqrt{\frac{r}{p}}$ with $\delta=\frac{1}{2}$ in the best case scenario, under the condition $\frac{k_S}{k_L}\leq \frac{1}{54}$.
The assumption $\delta \leq \alpha$ is made to prevent the violation of Assumption \ref{spikyPOETnew} under the condition
$\lambda_r({L}^{*})>C_2 \frac{\psi}{\xi^2(T)}$ of Theorem \ref{thmMinetop} .

Assumption \ref{tails} is necessary to ensure that the large deviation theory
can be applied to $f_{ij}$, $\epsilon_{ij'}$ and $f_{ij}\epsilon_{ij'}$ for all $i\leq n$, $j\leq r$ and $j'\leq p$.
Assumption \ref{tails2} is necessary to apply the results of \cite{bickel2008covariance} on the thresholding of the sparse component,
which prescribe that $S^*$ must be well conditioned with uniformly bounded diagonal elements.
We stress that the maximum degree $s'$ must be bounded. 
This condition is stronger than the corresponding one in \cite{fan2013large}, which prescribes $\max_{i\leq p} \sum_{j \leq p} |s^*_{ij}|^q~<~ c_4$, $q\in[0,1]$, $c_4>0$.
%
This is the price to pay for algebraic consistency, because our assumption ensures $\mu(\Omega({S^*}))=k_S p^{\delta}$ with $\delta \leq 1/2$.

Assumption \ref{pr} prescribes that the latent rank is infinitesimal with respect to $p$ and the sample size $n$ is possibly smaller than $p$,
but not smaller than $\delta_4 p^{\frac{3}{2}\delta}$. The need for this assumption rises throughout the proof of (\ref{samplePOET}), and to ensure consistency with Assumption \ref{tails2}. In fact, from the condition $S_{min,off}>C_3\frac{\psi}{\mu(\Omega)}$ of Theorem \ref{thmMinetop} it descends 
\begin{equation}S_{min,off}\times s' < \max \sum_{j \leq p} |s^*_{ij}| \leq \delta_2 p^{\delta}.\label{inS}\end{equation}
The inequality (\ref{inS}), under Assumptions \ref{alg}, \ref{tails2} and \ref{pr}, boils down to 
$\frac{1}{k_L \delta_4}p^{\alpha-2\delta}<\delta_2p^{\delta}$, which holds if Assumption \ref{sc} is respected.
As a consequence, we can allow for $||{S}^*||_2 \leq \max \sum_{j \leq p} |s^*_{ij}|\leq \delta_2 p^{\delta}$, $||{S}^*||_1\leq p \max \sum_{j \leq p} |s^*_{ij}| \leq \delta_2 p^{1+\delta}$ and $||{S}^*||_0=p+s\leq p s'\leq \delta_2 p^{1+\delta}$.




All outlined propositions must hold for finite values of $p$, $\alpha$ and $n$.
The following theorem provides a non-asymptotic consistency result
particularly useful when $p$ is not that large and $\alpha<1$,
because the absolute rate of ${\Sigma}_n$ under POET assumptions, $O(\frac{p}{\sqrt{n}})$,
may be too strong and prevent consistency.
\begin{thm}\label{thmMinetop}
Let $\Omega=\Omega({S}^{*})$ and
$T=T({L}^{*})$. Suppose that Assumptions \ref{spikyPOETnew}-\ref{sc} hold.
Define
$$
\psi=\frac{1}{\xi(T)}\frac{p^{\alpha}}{\sqrt{n}}
$$
with $\rho=\gamma \psi$, where $\gamma \in [9\xi(T),1/(6\mu(\Omega))]$.
In addition, suppose that
the minimum singular value of ${L}^{*}$ ($\lambda_r({L}^{*})$) is greater than
$C_2 \frac{\psi}{\xi^2(T)}$ and 
the smaller absolute value of the nonzero entries of ${S}^{*}$,
$S_{min,off}$, is greater than $C_3\frac{\psi}{\mu(\Omega)}.$ 
Then, with probability greater than
$1-C_4p^{-C_5}$, 
the pair $(\hat{{L}},\hat{{S}})$ minimizing (\ref{func:ob1})
recovers the rank of ${L}^{*}$ and the sparsity pattern of ${S}^{*}$ exactly:
$$rank(\hat{{L}})=rank({L}^{*})\:
\mbox{and}\: sign(\hat{{S}})=sign({S}^{*}).$$
Moreover, with probability greater than
$1-C_4p^{-C_5}$, the matrix losses for each component are bounded as follows:\\
$$||\hat{{L}}-{L}^{*}||_2\leq
C\psi, \qquad ||\hat{{S}}-{S}^{*}||_{\infty}\leq C\rho.$$
\end{thm}
We call the resulting covariance estimator ALCE (ALgebraic Covariance Estimator): $\hat{{\Sigma}}_{ALCE}=\hat{{L}}_{ALCE}+\hat{{S}}_{ALCE}$ .
The proof is reported in the Supplementary material. The technical key lies in proving the bound (\ref{samplePOET}).
The Theorem states that under all prescribed assumptions
the pair $(\hat{{L}},\hat{{S}})$ minimizing (\ref{func:ob1})
recovers exactly the rank of ${L}^{*}$ and the sparsity pattern of ${S}^{*}$,
provided that the minimum latent eigenvalue and the minimum residual absolute off-diagonal entry are large enough, as well as the
underlying matrix varieties $T$ and $\Omega$ are transverse enough.

We stress that the conditions $\lambda_r({L}^{*})>C_2 \frac{\psi}{\xi^2(T)}$ and $S_{min,off}>C_2\frac{\psi}{\mu(\Omega)}$ under Assumptions \ref{alg} and \ref{pr}
become $\lambda_r({L}^{*})>C_2 p^{\alpha}$ and $S_{min,off}> C_3 p^{\alpha-2\delta}$
respectively.
The latter in turn leads to (\ref{inS}) under Assumption \ref{sc}. Therefore, the resultant model setting is fully consistent with Assumptions \ref{spikyPOETnew} and \ref{tails2}.





Our results are non-asymptotic in nature, thus giving some probabilistic guarantees for finite values of $p$ and $n$. This fact depends on the algebraic consistency properties, which ensure the exact recovery of the rank and the sparsity pattern, with some parametric guarantees for the estimation error in $g_\gamma$ norm (see (\ref{tiprego})).
The shape of the probabilistic bound $\psi$ depends on the assumption $n\geq \delta_4 p^{3\delta}$,
which is also necessary in order to ensure asymptotic consistency, as
the following Corollary shows.
\begin{coroll}\label{asy}
Suppose that Assumptions \ref{spikyPOETnew}-\ref{pr} hold.\\
If the simultaneous limit $\lim_{\nu \rightarrow \infty}\min_{\nu}(p_\nu^{\alpha},n_\nu)=\infty$
with the path-wise restriction\\ $\lim_{\nu \rightarrow \infty} \frac{p_\nu^{2\alpha}}{n_\nu}=0$ holds, then
$\psi=\left(\frac{1}{\xi(T)}\frac{p^{\alpha}}{\sqrt{n}}\right)$ tends to $0$. 
\end{coroll}
Corollary \ref{asy} states the asymptotic consistency of the estimates, showing how the probabilistic error annihilates.
For the terminology about limit sequences see \cite{bai2003inferential}.
Moreover, $\frac{\psi}{p^{\alpha}}\rightarrow 0$ as $\lim_{\nu \rightarrow \infty}\min_{\nu}(p_\nu^{\alpha},n_\nu)=\infty$, thus establishing the asymptotic consistency in
relative terms even if $\alpha<1$, resembling the ''blessing of dimensionality'' described in \cite{fan2013large}.

We stress that the probabilistic bound $\psi$ decreases to $0$ as long as $n \geq \delta_4 p^{\frac{3}{2}\delta}$.
This assumption leads to overcome the restrictive condition $p \leq n$, since $\delta \leq 1/2$.
In addition, 
an immediate consequence of (\ref{samplePOET}) is reported in the following Corollary (the proof is reported in the Supplement).
\begin{coroll}\label{rc}
Let $\hat{\lambda}_{r,ALCE}$ be the $r-$th largest eigenvalue of $\hat{{L}}_{ALCE}$. Then under the assumptions of Theorem \ref{thmMinetop} $\hat{\lambda}_{r,ALCE}>C_1 p^{\alpha}$ with probability approaching $1$ for some $C_1>0$.
\end{coroll}
Corollary \ref{rc} states that $\hat{{L}}_{UNALCE}$ is rank-consistent as the latent degree of spikiness $\alpha$ is not smaller than the maximum degree of the residual component $\delta$.
If $\alpha=1$, we fall back to the POET setting. 
If $\alpha=0$, $r=\log(1)=0$, and we fall back to the pure sparsity estimator of \cite{bickel2008covariance}.

A representative selection of the latent eigenvalue and sparsity patterns admitted under described conditions is reported in the Supplementary Material.
We emphasize that the algebraic consistency does no longer force the latent eigenvalues to scale to $p$, provided that the spectral norm of the residual component is scaled accordingly. In general, it is needed that the minimum latent eigenvalue and absolute nonzero residual entry are large enough to ensure consistency, but, unlike POET, they can be scaled to $p^{\alpha}$, $\alpha<1$. The exponent $\alpha$ plays the role of an adaptive spikiness degree.

In particular, if we increase $\alpha$ (\emph{ceteris paribus}), both $\lambda_r({L}^{*})$ and $S_{min,off}$ must be larger to ensure identifiability.
The same happens if $p$ increases, because, according to \cite{chandrasekaran2011rank},
both $\xi(T({L}^{*}))$ and $\mu(\Omega({S}^{*}))$ depend inversely on $p$.
On the contrary, to ensure consistency, if $r$ increases ${L}^{*}$
can have less spiked eigenvalues, while if $s'$ increases ${S}_{min,off}$ can be smaller.
This occurs because, according to \cite{chandrasekaran2011rank}, $\xi(T({L}^{*}))$ and $\mu(\Omega({S}^{*}))$ directly depend on $r$ and $s'$ respectively.




From Theorem \ref{thmMinetop},
we can derive with probability larger than $1-C_1p^{-C_2}$ the following bounds for $\hat{{\Sigma}}_{ALCE}$:
\begin{eqnarray}
||\hat{{\Sigma}}_{ALCE}-{\Sigma}^{*}||_{2} & \leq & C(s'\xi(T)+1)\psi=\phi,\\
||\hat{{\Sigma}}_{ALCE}-{\Sigma}^{*}||_{Fro} & \leq & C(\sqrt{ps'}\xi(T)+\sqrt{r}) \psi,
\end{eqnarray}
which hold if and only if $\lambda_{min}({\Sigma}^{*}) > \phi$.
The same bounds hold for the inverse covariance estimate $\hat{{\Sigma}}^{-1}_{ALCE}$ with the same probability:
\begin{eqnarray} 
||\hat{{\Sigma}}^{-1}_{ALCE}-{\Sigma}^{*-1}||_{2} & \leq & C(s'\xi(T)+1)\psi=\phi\\
||\hat{{\Sigma}}^{-1}_{ALCE}-{\Sigma}^{*-1}||_{Fro} & \leq & C(\sqrt{ps'}\xi(T)+\sqrt{r}) \psi
\end{eqnarray}
given that $\lambda_{min}({\Sigma}^{*}) \geq 2\phi$.

Within the same framework, we can complete our analysis with the bounds for $\hat{{S}}$.
From $||\hat{{S}}-{S}^{*}||\leq s'||\hat{{S}}-{S}^{*}||_\infty $, we obtain \begin{equation}||\hat{{S}}-{S}^{*}||_{2} \leq Cs'\xi(T)\psi=\phi_{{S}}.\end{equation}
From $||\hat{{S}}-{S}^{*}||_{Fro} \leq \sqrt{ps'}||\hat{{S}}-{S}^{*}||_{\infty}$, we obtain \begin{equation}||\hat{{S}}-{S}^{*}||_{Fro}\leq C\sqrt{ps'}\xi(T)\psi.\end{equation}
$\hat{{S}}$ is positive definite if and only if $\lambda_{min}({{S}^{*}})>\phi_{{S}}.$
$\hat{{S}}^{-1}$ has the same bound of $\hat{{S}}$ if and only if $\lambda_{min}({S}^{*})\geq 2\phi_{{S}}$.

To sum up, by ALCE estimator we offer the chance to recover consistently a relaxed spiked eigen-structure, thus
overcoming the condition $p\leq n$, even using the sample covariance matrix as estimation input (the ratio $p/n$  directly impacts on the error bound).
Our bounds are in absolute norms, and reflect the underlying degree of spikiness $\alpha$.
%
%
%
Our theory relies on the probabilistic convergence of the sample covariance matrix under the assumption that the data follow an approximate factor model with a sparse residual.
If $p$ and $n$ are in a proper relationship, both parametric and algebraic consistency are ensured.


\subsection{UNALCE estimator: a re-optimized ALCE solution}
Let us define ${\Delta}_{{L}}=\hat{{L}}_{ALCE}-{L}^{*}$,${\Delta}_{{S}}=\hat{{S}}_{ALCE}-{S}^{*}$,${\Delta}_{{\Sigma}}=\hat{{\Sigma}}_{ALCE}-{\Sigma}^{*}$.
A key aspect of Theorem \ref{thmMinetop} is that the two losses in ${L}^{*}$ and ${S}^{*}$ are bounded separately. This fact results in a negative effect on the overall performance of $\hat{{\Sigma}}_{ALCE}$, represented by the loss $||{\Delta}_{{\Sigma}}||_{2}$,
since $||{\Delta}_{{\Sigma}}||_2$ is simply derived as a function of $||{\Delta}_{{L}}||_2$ and $||{\Delta}_{{S}}||_2$ according to the triangle inequality $||{\Delta}_{{\Sigma}}||_2\leq ||{\Delta}_{{L}}||_2+ ||{\Delta}_{{S}}||_2$. 
Therefore, the need rises to correct for this drawback, re-shaping $\hat{{\Sigma}}_{ALCE}$, as ALCE approach is somehow
sub-optimal for the whole covariance matrix. 


We approach this problem  by a finite-sample analysis, which could be referred to as a re-optimized least squares method.
We refer to the usual objective function (\ref{func:ob1}) with $||{S}||_1=||{S}||_{1,off}=\sum_{i=1}^{p-1}\sum_{j=i+1}^p |s_{ij}|$, i.e. the $l_1$ norm of ${S}$ excluding the diagonal entries, consistently with POET approach.
We define ${Y}_{pre}$ and ${Z}_{pre}$ the last updates in the gradient step of the minimization algorithm of (\ref{func:ob1}) (see the Supplement for more details).
${Y}_{pre}$ and ${Z}_{pre}$ are the two matrices we condition upon in order to derive our finite-sample re-optimized estimates.


Suppose that $\hat{\mathcal{B}}(\hat{r})$
and $\hat{\mathcal{A}}(\hat{s})$
are the recovered varieties ensuring the algebraic consistency of (\ref{func:ob1}).
One might look for the solution (say
$(\hat{{L}}_{New},\hat{{S}}_{New})$) of the problem
\begin{equation}
\min_{{L} \in \hat{\mathcal{B}}(\hat{r}), {S}\in
\hat{\mathcal{A}}(\hat{s})} TL({L},{S})=||({\Sigma}_n-({L}+{S})||^2_{Fro}\label{TL},
\end{equation}
where $TL({L},{S})$ stands for \emph{Total Loss}.
The sample covariance matrix follows the model ${\Sigma}_n={L}^{*}+{S}^{*}+{W}$,
given a sample of $p-$dimensional data vectors ${x}_i$, $i=1,\ldots,n$.
Our problem essentially is: which pair
${L} \in \hat{\mathcal{B}}(\hat{r}), {S}\in
\hat{\mathcal{A}}(\hat{s})$ satisfying algebraic consistency shows the best approximation properties of ${\Sigma}_n$?

We prove the following result.
\begin{thm}\label{mine}
Suppose that $\hat{\mathcal{B}}(r)$, and $\hat{\mathcal{A}}(s)$ are the recovered matrix varieties, and that $\hat{{L}}_{ALCE}=\hat{ U}_{ALCE}\hat{ D}_{ALCE}\hat{ U}_{ALCE}'$ is the eigenvalue decomposition of $\hat{{L}}_{ALCE}$.
Define $\hat{{S}}_{New}$ such that its off-diagonal elements are the same as $\hat{{S}}_{ALCE}$ and $\hat{{\Sigma}}_{New}$ such that its diagonal elements are the same as $\hat{{\Sigma}}_{ALCE}$ respectively.
Then, the minimum
\begin{equation}\min_{{L}\in\hat{\mathcal{B}}(\hat{r}), {S} \in \hat{\mathcal{A}}(\hat{s})}\|{\Sigma}_{n}-({L}+{S})\|^2_{Fro} \label{min_opt}\end{equation}
conditioning on ${Y}_{pre}$ and ${Z}_{pre}$ is achieved if and only if
\[\hat{{L}}_{new}=\hat{ U}_{ALCE}(\hat{ D}_{ALCE}+\breve{\psi} {I}_r)\hat{ U}_{ALCE}' \quad  \mbox{   and if   } \quad
 diag(\hat{{S}}_{New})=diag(\hat{{\Sigma}}_{ALCE})-diag(\hat{{L}}_{new})\] where $\breve{\psi}>0$ is any prescribed threshold parameter. 
\end{thm}

Theorem \ref{mine} essentially states that the sample total loss (\ref{TL}) is minimized if we un-shrink the eigenvalues of $\hat{{L}}_{ALCE}$ (re-adding the threshold $\breve{\psi}$). We call the resulting overall estimator $\hat{{\Sigma}}_{new}=\hat{{L}}_{new}+\hat{{S}}_{new}$ UNALCE (UNshrunk ALgebraic Covariance Estimator).  We stress the importance of conditioning on ${Y}_{pre}$ and ${Z}_{pre}$.
Since ${Y}_{pre}$ and ${Z}_{pre}$ are the matrices minimizing $\frac{1}{2}||({L}+{S})-{\Sigma}_n||_{Fro}^2$
conditioning on the contemporaneous minimization of $\breve{\psi}
||{L}||_{*} + \breve{\rho} ||{S}||_{1}$, our finite-sample re-optimization step aims to re-compute $\min||{\Sigma}_{n}-({L}+{S})||^2_{Fro}$
once removed the effect of the composite penalty.

As proved in the Supplement (which we refer to for the details), problem (\ref{TL}) can be decomposed in two problems: one in ${L}$ and one in ${S}$.
The problem in ${L}$ is solved by the covariance matrix formed by the top $\hat{r}$ principal components of $ {Y}_{pre}$, which belongs by construction to $\hat{\mathcal{B}}(\hat{r})$ and is equal to $\hat{ U}_{ALCE}(\hat{ D}_{ALCE}+\breve{\psi} {I}_r)\hat{ U}_{ALCE}'=\hat{{L}}_{UNALCE}$.
The problem in ${S}$ collapses to the problem in ${L}$ under the prescribed assumptions on the off-diagonal elements of $\hat{{S}}_{UNALCE}$ (which causes $\hat{{S}}_{UNALCE} \in \hat{\mathcal{A}}(\hat{s})$), and on the diagonal elements of $\hat{{\Sigma}}_{UNALCE}$. 
The new estimate of the diagonal of ${S}^*$ is simply the difference between the diagonal of the original $\hat{{\Sigma}}_{ALCE}$
and the diagonal of the newly computed $\hat{{L}}_{UNALCE}$.
Note that our re-optimization step depends entirely on ${\Sigma}_n$, as ${Y}_{pre}$ and ${Z}_{pre}$ are ${\Sigma}_n$-dependent.


Four consequences of Theorem \ref{mine} are reported in Corollary \ref{mine2}.
\begin{coroll}\label{mine2}
The gains in terms of spectral loss for $\hat{{L}}_{UNALCE}$, $\hat{{S}}_{UNALCE}$ in comparison to $\hat{{L}}_{ALCE}$, $\hat{{S}}_{ALCE}$ respectively are all strictly positive and bounded by $\breve{\psi}$:
\begin{eqnarray}
0<||\hat{{L}}_{ALCE}-{L}^{*}||_2-||\hat{{L}}_{UNALCE}-{L}^{*}||_2 & \leq & \breve{\psi},\\
0<||\hat{{S}}_{ALCE}-{S}^{*}||_{2}-||\hat{{S}}_{UNALCE}-{S}^{*}||_{2} & \leq &\breve{\psi}.
\end{eqnarray}
The gains in terms of Frobenius norm are all strictly positive and bounded as follows:
\begin{eqnarray}
0<||\hat{{L}}_{ALCE}-{L}^{*}||_{Fro}-||\hat{{L}}_{UNALCE}-{L}^{*}||_{Fro} & \leq & \sqrt{r}\breve{\psi},\\
0<||\hat{{S}}_{ALCE}-{S}^{*}||_{Fro}-||\hat{{S}}_{UNALCE}-{S}^{*}||_{Fro} & \leq & \sqrt{r} \breve{\psi}.
\end{eqnarray}
\end{coroll}

Two further relevant consequences of Theorem \ref{mine} are reported in Corollary \ref{mine3}.
\begin{coroll}\label{mine3}
The gain in terms of spectral sample total loss for $\hat{{\Sigma}}_{UNALCE}$ respect to $\hat{{\Sigma}}_{ALCE}$ is strictly positive and bounded by $\breve{\psi}$:
\begin{equation}
0<||{\Sigma}_n-\hat{{\Sigma}}_{ALCE}||_{2}-||{\Sigma}_n-\hat{{\Sigma}}_{UNALCE}||_{2}\leq \breve{\psi}.
\end{equation}
The gain in terms of Frobenius sample total loss for $\hat{{\Sigma}}_{UNALCE}$ respect to $\hat{{\Sigma}}_{ALCE}$
is strictly positive and bounded by $\sqrt{r}\breve{\psi}$:
\begin{equation}
0<||{\Sigma}_n-\hat{{\Sigma}}_{ALCE}||_{Fro}-||{\Sigma}_n-\hat{{\Sigma}}_{UNALCE}||_{Fro}\leq \sqrt{r}\breve{\psi}.
\end{equation}
\end{coroll}





The following result compares the losses of $\hat{{\Sigma}}_{UNALCE}$ and $\hat{{\Sigma}}_{ALCE}$ from the target ${\Sigma}^{*}$.
\begin{thm}\label{mine4}
Conditioning on ${\Sigma}_n$, the gains in terms of spectral loss and Frobenius loss for $\hat{{\Sigma}}_{UNALCE}$ respect to $\hat{{\Sigma}}_{ALCE}$
are strictly positive and bounded as follows:
\begin{eqnarray}
0 < ||\hat{{\Sigma}}_{ALCE}-{\Sigma}^{*}||_{2}-||\hat{{\Sigma}}_{UNALCE}-{\Sigma}^{*}|||_{2} & \leq & \breve{\psi},\\
0 < ||\hat{{\Sigma}}_{ALCE}-{\Sigma}^{*}||_{Fro}-||\hat{{\Sigma}}_{UNALCE}-{\Sigma}^{*}|||_{Fro} & \leq & \sqrt{r}\breve{\psi}.
\end{eqnarray}
\end{thm}
 The rationale of the reported claims is the following.
 We accept to pay the price of a non-optimal solution in terms of nuclear norm (we allow to increment $||\hat{{L}}||_{*}$ by $r \breve{\psi}$) but we have a best fitting performance for the whole covariance matrix, decrementing the squared Frobenius loss of $\hat{{\Sigma}}$ by a quantity bounded by $r \breve{\psi}^2$.
 The $l_1$ norm of ${S}$ excluding the diagonal, $||\hat{{S}}||_{off}$, is unvaried, while the norm $||{S}||_1$ (included the diagonal)
 is decreased by a quantity bounded by $\sqrt{r} \breve{\psi}$.

The following Corollary extends our framework to the performance of $(\hat{{\Sigma}}_{UNALCE})^{-1}$.
\begin{coroll}\label{mine5}
The gains in terms of spectral loss and Frobenius loss for $\hat{{\Sigma}}_{UNALCE}^{-1}$ respect to $\hat{{\Sigma}}_{ALCE}^{-1}$ are strictly positive and bounded as follows:
\begin{eqnarray}0<||\hat{{\Sigma}}_{ALCE}^{-1}-{\Sigma}^{*-1}||_{2}-||(\hat{{\Sigma}}_{UNALCE}^{-1}-{\Sigma}^{*-1}||_{2} & \leq & \breve{\psi}.\\
0<||\hat{{\Sigma}}_{ALCE}^{-1}-{\Sigma}^{*-1}||_{Fro}-||\hat{{\Sigma}}_{UNALCE}^{-1}-{\Sigma}^{*-1}||_{Fro} & \leq & \sqrt{r}\breve{\psi}.
\end{eqnarray}
\end{coroll}


The outlined results allow to improve the estimation performance given the finite sample.
However, the non-asymptotic bounds for $\hat{{L}}_{UNALCE}$, $\hat{{S}}_{UNALCE}$ and $\hat{{\Sigma}}_{UNALCE}$ are exactly the ones of $\hat{{L}}_{ALCE},\hat{{S}}_{ALCE}$ and $\hat{{\Sigma}}_{ALCE}$. UNALCE improves systematically the fitting performance of ALCE, inheriting all its algebraic and parametric consistency properties. The proofs of all theorems and corollaries can be found in the Supplement.

Finally, we study how the necessary conditions to ensure the positive definiteness of UNALCE estimates evolve respect to the ALCE ones.
The following Corollary holds.
\begin{coroll}\label{defin}
$\hat{L}_{UNALCE}$ is positive semi-definite if ${\lambda}_r({{L}^{*}}) \geq \delta_{\alpha} p^{\alpha} -\breve{\psi}$.
$\hat{S}_{UNALCE}$ is positive definite if ${\lambda}_p({{S}^{*}}) > \phi_S+\frac{r}{p} \breve{\psi}$.
$\hat{\Sigma}_{UNALCE}$ is positive definite if ${\lambda}_p({{\Sigma}^{*}}) > \phi+\frac{r}{p} \breve{\psi}$.
\end{coroll}
We stress that the improvement of the condition for $\hat{{L}}_{UNALCE}$ is numerically much larger than the worsening of the conditions
for ${\hat{S}}_{UNALCE}$ and  ${\hat{\Sigma}}_{UNALCE}$.


\section{A new model selection criterion: $MC$}\label{crit}

In empirical applications, the selection of thresholds $\psi$ and $\rho$ in equation (\ref{func:ob1}) requires a model selection criterion
consistent with the described estimation method.
The motivation rises from the consistency norm $g_\gamma$ used in \cite{luo2011recovering} (see (\ref{ggamma2})). 
Our aim is to detect the optimal threshold pair $(\psi,\rho)$ in respect to the spikiness/sparsity trade-off.
In order to exploit (\ref{ggamma2}) with model selection purposes, we need to make the two terms comparable, i.e.,
the need of rescaling both arguments of $g_\gamma$ rises.


First of all, we note that if all the estimated latent eigenvalues are equal, we have $||\hat{{L}}||_{*}=\hat{r}||\hat{{L}}||$.
As the condition number of $\hat{{L}}$ increases, we have $\hat{r}||\hat{{L}}||>||\hat{{L}}||_{*}$.
As a consequence, the quantity $\hat{r}||\hat{{L}}||$
acts as a penalization term against the presence of too small eigenvalues.
Analogously, if $\hat{{S}}$ is diagonal it holds ${||\hat{{S}}||_{\infty}}= {||\hat{{S}}||_{1,v}}$.
As the number of non-zeros increases, it holds ${||\hat{{S}}||_{1,v}}>{||\hat{{S}}||_{\infty}}$.
Therefore, the quantity ${||\hat{{S}}||_{1,v}}$ acts as a penalization term against the presence of too many non-zeros.

In order to compare the magnitude of the two quantities, we divide the former
by the trace of $\hat{{L}}$, estimated by
$\hat{\theta}trace({\Sigma}_n)$, and the latter by the trace of $\hat{{S}}$, estimated by $(1-\hat{\theta})trace({\Sigma}_n)$.
Our maximum criterion $MC$ can be therefore defined as follows:
\begin{equation}MC(\psi,\rho)=\max
\left\{\frac{\hat{r}||\hat{{L}}||_2}{\hat{\theta}},
\frac{{||\hat{{S}}||_{1,v}}}{{\gamma}(1-\hat{\theta})}\right\},
\end{equation}
where ${\gamma}=\frac{\rho}{\psi}$
is the ratio between the sparsity and the spikiness threshold.

MC criterion is by definition mainly intended to catch the proportion of variance explained by the factors.
For this reason, it tends to choose quite sparse solutions with a small number of non zeros and a small proportion of residual covariance,
unless the non-zero entries of $\hat{{S}}$ are prominent, as Theorem \ref{thmMinetop} prescribes.
The $MC$ method performs considerably better than the usual cross-validation using $H$-fold Frobenius loss (used in \cite{luo2011recovering}).
In fact, minimizing a loss based on a sample approximation like the Frobenius one causes the parameter $\hat{\theta}$ to be shrunk too much.
The threshold setting which shows a minimum
for $MC$ criterion (given that the estimate $\hat{{\Sigma}}$ is positive definite) is the best
in terms of composite penalty, taking into account the latent low rank and sparse structure simultaneously.

\section{A Euro Area banking data example}\label{real}

This Section provides a real example on the performance of POET and UNALCE based on a selection of Euro Area banking data.
We acknowledge the assistance of the European Central Bank, where one of the authors spent a semester as a PhD trainee,
in providing access to high-level banking data.
Here we use the covariance matrix computed on a selection of balance sheet indicators for some of the most relevant Euro Area banks by systemic power.
The overall number of banks (our sample size) is $n=365$.
These indicators are the ones needed for supervisory reporting, and include capital and financial variables.

\begin{figure}[htb]
\centerline{\,
\includegraphics[width=0.6\textwidth]{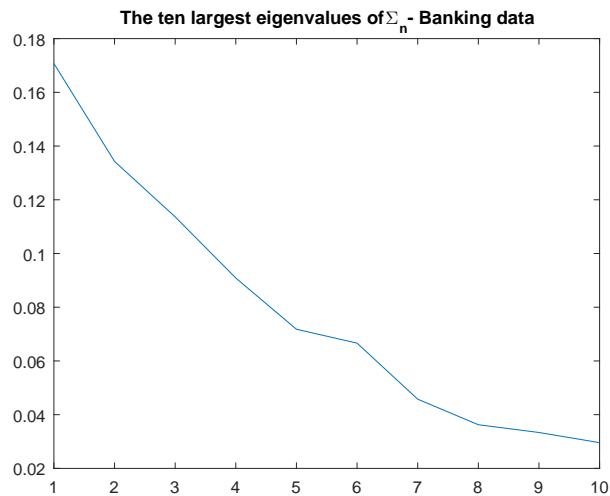}}
\caption{Supervisory data: sample eigenvalues}
\label{eigECB}
\end{figure}

The chosen raw variables ($1039$) were rescaled to the total asset of each bank.
Then, a screening based on the importance of each variable, intended as the absolute amount of correlation with all the other variables,
was performed in order to remove identities.
The resulting very sparse data matrix contains $p=382$ variables: here we are in the typical $p>n$ case, where the sample covariance matrix is completely ineffective.
We plot sample eigenvalues in Figure \ref{eigECB}.

UNALCE estimation method selects a solution having a latent rank equal to $6$.
The number of surviving non-zeros in the sparse component is $328$, which is the $0.45\%$ of $72772$ elements.
Conditioning properties are inevitably very bad. The results are reported in Table \ref{T1ECB}.


\begin{table}
\caption{\label{T1ECB}Supervisory data: results for $\hat\Sigma_{UNALCE}$}
\centering
\fbox{%
\begin{tabular}{cc}
\em \textbf{Supervisory data} & \em UNALCE \\
  $\hat{r}$ & 6 \\
  nz & 328 \\
  $perc_{nz}$ & 0.0045 \\
  $\hat{\theta}$ &  0.3247\\
  $\hat{\rho}_{corr}$ & 0.1687 \\
  Sample TL & $0.0337$ \\
  $cond(\hat{{\Sigma}})$ & 6.35E+15 \\
  $cond(\hat{{S}})$ & 2.78E+15 \\
  $cond(\hat{{L}})$ & 3.1335 \\
\end{tabular}}
\end{table}

\begin{table}
\caption{\label{T2ECB}Supervisory data: results for $\hat\Sigma_{POET}$}
\centering
\fbox{%
\begin{tabular}{cc}
\em \textbf{Supervisory data} & \em POET \\
  $\hat{r}$ & 6 \\
  nz & 404 \\
  $perc_{nz}$ & 0.0056 \\
  $\hat{\theta}$ & 0.6123 \\
  $\hat{\rho}_{corr}$ & 0.0161 \\
  Sample TL & $0.0645$ \\
  $cond(\hat{{\Sigma}})$ & 6.68E+15 \\
  $cond(\hat{{S}})$ & 1.11E+15 \\
  $cond(\hat{{L}})$ & 2.5625 \\
\end{tabular}}
\end{table}




In order to to obtain
a POET estimate, we exploit the algebraic consistency of $\hat\Sigma_{UNALCE}$
setting the rank to $6$ and we perform cross-validation for threshold selection.
The results are reported in Table \ref{T2ECB}, where we note
that the number of estimated non-zeros is $404$ ($0.56\%$).


Apparently, one could argue that POET estimate is better: the estimated proportion of common variance is $0.6123$,
and the proportion of residual covariance is $0.0161$. On the contrary, UNALCE method outputs $\hat{\theta}=0.3247$
and $\hat{\rho}_{corr}= 0.1687$.
A relevant question arises: how much is the true proportion of variance explained by the factors?
In fact, a so high latent proportion variance, which depends on the use of PCA with $6$ components,
causes the residual covariance proportion to be very low.
Therefore, POET procedure gives \emph{a priori} a preference for the low rank part.
This pattern does not change even if we choose a lower value for the rank.

On the contrary, the UNALCE estimate, which depends on a double-step iterative thresholding procedure,
requires a larger magnitude of the non-zero elements in the sparse component.
In fact, the proportion of lost covariance during the procedure is here $29.39\%$.
As a consequence, via rank/sparsity detection UNALCE shows better approximation properties respect to POET:
its Sample Total Loss is relevantly lower than the one of the competitor ($0.337$ VS $0.645$).

For UNALCE method, the covariance structure appears so complex that a relevant proportion
of residual covariance is present.
This allows us to explore the importance of variables, that is to explore which variables have the largest systemic power
(i.e. the most relevant communality) or the largest idiosyncrasy (i.e. the most relevant residual variance).

\begin{figure}[htb]
\centerline{\,
\includegraphics[width=0.7\textwidth]{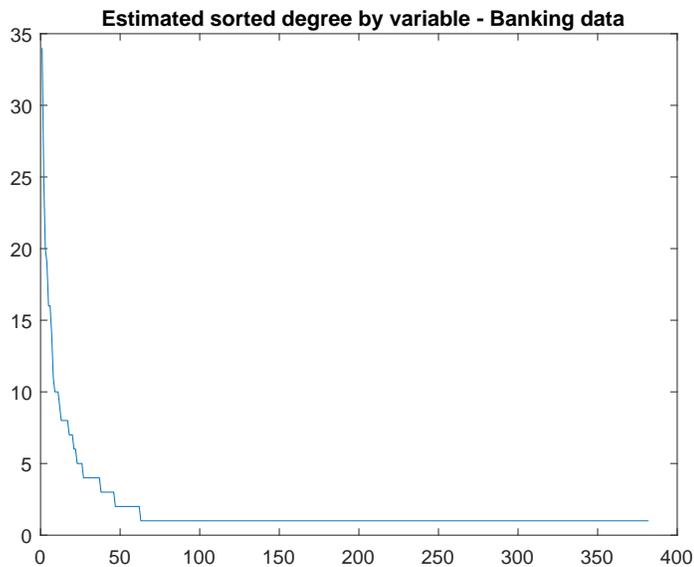}}
\caption{Banking data: sorted degree by variable}
\label{Deg_rank}
\end{figure}

In Figure \ref{Deg_rank} we plot the estimated degree (number of non-zero covariances in the residual component)
sorted by variable. Only $62$ out of $382$ variables have at least one non-zero residual covariance.

\begin{figure}[htb]
\centerline{\,
\includegraphics[width=1\textwidth]{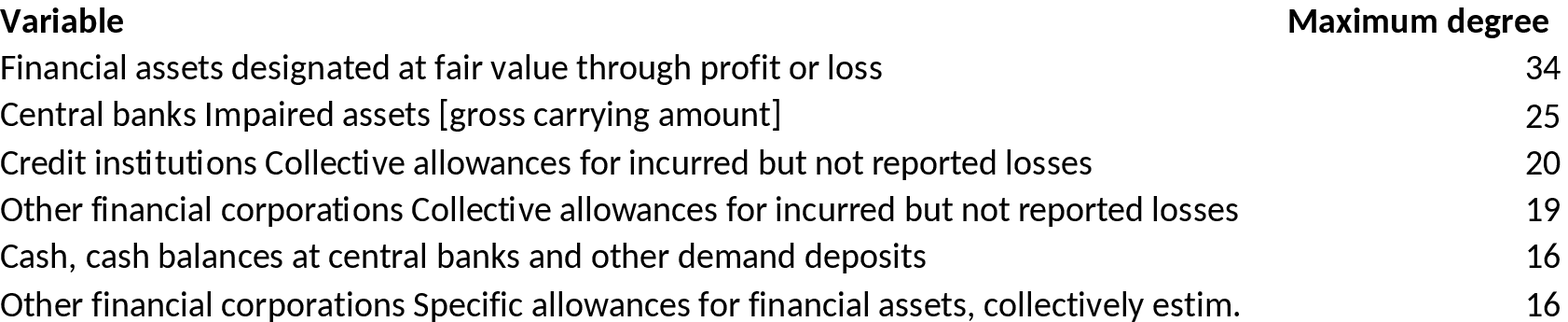}}
\caption{Banking data: top $6$ variables by degree}
\label{Deg_def}
\end{figure}

In Figure \ref{Deg_def} we report the top 6 variables by estimated number of non-zero residual covariances. They are mainly credit-based variables: financial assets through profit and loss,
central banks impaired assets, allowances to credit institutions and non-financial corporations, cash.
These variables are related to the largest number of other variables.

\begin{figure}[htb]
\centerline{\,
\includegraphics[width=1\textwidth]{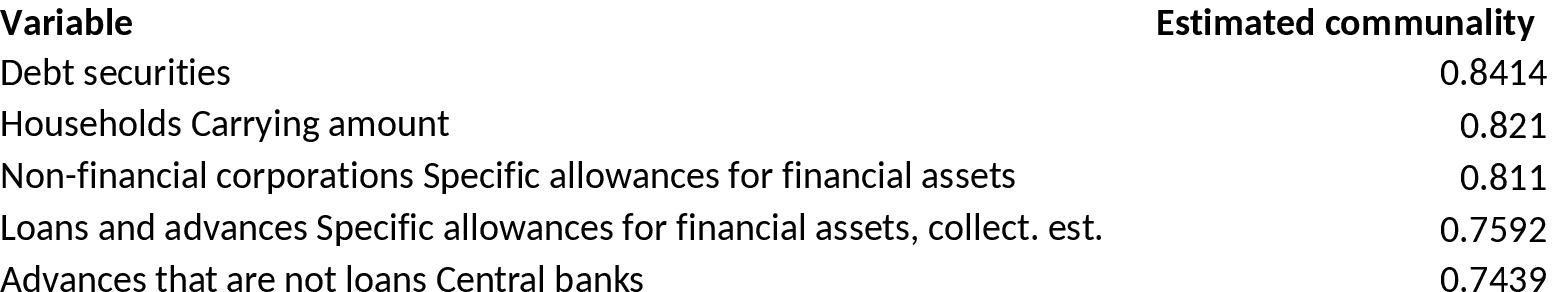}}
\caption{Banking data: top $5$ variables by estimated communality}
\label{ECB_comm}
\end{figure}

In Figure \ref{ECB_comm} we report the top 5 variables by estimated communality, defined as
$$\frac{\hat{l}_{UNALCE,jj}}{\hat{\sigma}_{UNALCE,jj}} \: \forall j=1,\ldots,382.$$
The results are very meaningful: the most systemic variables are debt securities, loans and advances to households, specific allowances for financial assets,
advances which are not loans to central banks. All these are fundamental variables for banking supervision, because they represent key indicators
for the assessment of bank performance.

\begin{figure}[htb]
\centerline{\,
\includegraphics[width=1\textwidth]{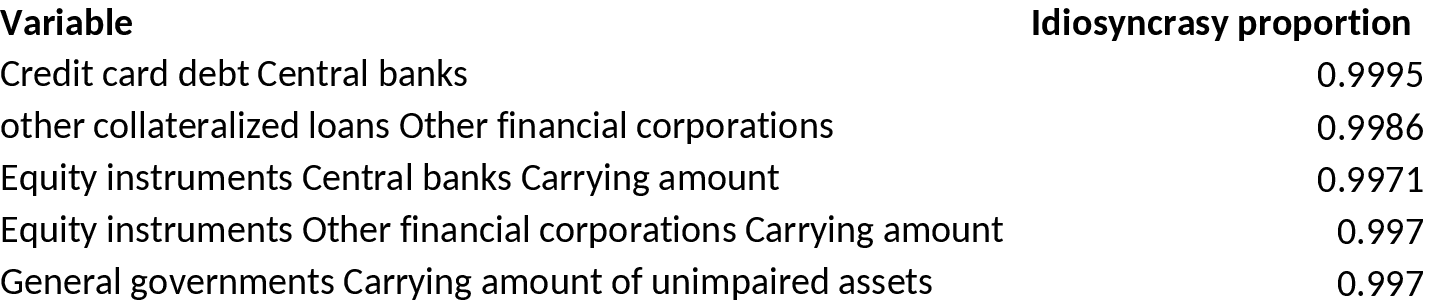}}
\caption{Banking data: top $5$ variables by residual covariance proportion}
\label{ECB_res}
\end{figure}

In Figure \ref{ECB_res} we report the top 5 variables by estimated idiosyncratic covariance proportion
$$\frac{\hat{s}_{UNALCE,jj}}{\hat{\sigma}_{UNALCE,jj}} \: \forall j=1,\ldots,382.$$
We note that those variables have a marginal power in the explanation of the common covariance structure,
and are much less relevant for supervisory analysis than the previous five.

In conclusion, our UNALCE procedure offers a more realistic view of the underlying covariance structure of a set of variables,
allowing a larger part of covariance to be explained by the residual sparse component respect to POET. 

\section{Conclusions}\label{concl}

The present work describes a numerical estimator of large covariance matrices which are assumed to be the sum of a low rank and a sparse component.
Estimation is performed solving a regularization problem where the objective function is composed by a smooth Frobenius loss and a non smooth composite penalty,
which is the sum of the nuclear norm of the low rank component and the $l_1$ norm of the sparse component.
Our estimator is called UNALCE (UNshrunk ALgebraic Covariance Estimator) and provides consistent recovery of the low rank and the sparse component, as well as of the overall covariance matrix, under a generalized assumption of spikiness of the latent eigenvalues.

In this paper we compare UNALCE and POET (Principal Orthogonal complEment Thresholding, \cite{fan2013large}), an asymptotic estimator which performs PCA to recover the low rank component and uses a thresholding algorithm to recover the sparse component. Both estimators provide the usual parametric consistency, while UNALCE provides also the algebraic consistency of the estimate, that is, the rank and the position of residual non-zeros are simultaneously detected by the solution algorithm.
This automatic recovery is a crucial advantage respect to POET: the latent rank, in fact, is automatically selected and the sparsity pattern of the residual component
is recovered considerably better.


In particular, we prove that UNALCE can effectively recover the covariance matrix even in presence of spiked eigenvalues with rate $O(p)$, exactly as POET estimator does, allowing $n$ to be as small as $O(p^{3 \delta})$, where $\delta$ is the maximum degree of the sparse component. In addition, we prove that the recovery is actually effective even if the latent eigenvalues show an intermediate degree of spikiness $\alpha \in [0,1]$. The resulting loss is bounded accordingly to $\alpha$ and the $r-$th latent eigenvalue is asymptotically strictly positive under the assumption $\delta \leq \alpha$. In this way we encompass both LOREC and POET theory in a generalized theory of large covariance matrix estimation by low rank plus sparse decomposition.



A real example on a set of Euro Area banking data shows that our tool is particularly useful
for mapping the covariance structure among variables even in a large dimensional context.
The variables having the largest systemic power, that is, the ones most affecting the common covariance structure, can be identified, as well as the variables having the largest idiosyncratic power, that is, the ones most characterized by the residual variance.
In addition, the variables showing the largest idiosyncratic covariances with all the other ones can be identified, thus recovering the strongest related variables.
Particular forms of the residual covariance pattern can thus be detected if present.

Our research may be ground for possible future developments in many directions.
In the time series context, this procedure can be potentially extended to covariance matrix estimation
under dynamic factor models.
Another fruitful extension of our procedure is related to the spectral matrix estimation context.
Finally, this tool can be potentially used in the Big data context, 
where both the dimension and the sample size are very large.
 This poses new computational and theoretical challenges,
 the solution of which is crucial to further extend the power of statistical modelling
 and its effectiveness in detecting patterns and underlying drivers of real phenomena.

\section*{Supplementary material}\label{concl}

The present paper is complemented by an Appendix containing a simulation study and the proofs of stated theorems and corollaries.
In addition, the MATLAB functions \verb|UNALCE.m| and \verb|POET.m|, performing UNALCE and POET procedures respectively, can be downloaded at 
\cite{dataset}.
Both functions contain the detailed explanation of input and output arguments.
Finally, the MATLAB dataset \verb|supervisory_data.mat|, which contains the covariance matrix, \verb|C|,
and the relative labels of supervisory indicators, \verb|Labgood|, can also be downloaded at the same link, which we refer to for the details.


\appendix

\section{A simulation study}\label{sim}

\subsection{Simulation settings}\label{simsett}

In order to compare
the performance of UNALCE, LOREC and POET,
we take into consideration five simulated low rank plus sparse settings reported in Table \ref{sett}.
The key simulation parameters are:
\begin{enumerate}
\item the dimension $p$, the sample size $n$;
\item the rank $r$ and the condition number $c$ of the low rank component ${L}^*$;
\item the trace of ${L}^*$, $\tau \theta p$, where $\tau$ is a magnitude parameter
and $\theta$ is the percentage of variance explained by ${L}^*$;
\item the (half) number non-zeros $s$ in the sparse component ${S}^*$;
\item the proportion of nonzeros $prop_s$;
\item the proportion of (absolute) residual covariance $\rho_{corr}$.
\item $N=100$ replicates for each setting.
\end{enumerate}

The reported settings give an exhaustive idea of the low rank plus sparse settings recoverable under our assumptions. 
The critical parameters are:
\begin{enumerate}
\item the spectral norm of ${L}^*$, which controls for the degree of spikiness. $||{L}^*||$ is a direct function of $\tau$ and an inverse function of $c$, which together control
for the magnitude of $\lambda_r({L}^*)$;
\item the spectral norm of ${S}^*$, which controls for the degree of sparsity. $||{S}^*||$ is a direct function of $prop_s$, which control for $s$ and $\rho_{corr}$. 
\end{enumerate}
All the norms relative to our simulated settings are reported in Table \ref{specond}.
The data generation algorithm is described in detail in \cite{farne2016large}.

\begin{table}
  \caption{\label{sett} Simulated settings: parameters}
  \centering
  \fbox{
  \begin{tabular}{cccccccccc|}
    \textbf{Setting}  & \em $p$ & \em $n$ & \em $r$ & \em $\theta$ & \em $c$ & \em $prop_s$ & \em $\rho_{corr}$\\
    \textbf{1} & $100$ & $1000$ & $4$ & $0.7$ & $2$ & $0.0238$ & $0.0045$\\
    \textbf{2} & $100$ & $1000$ & $4$ & $0.7$ & $4$ & $0.0677$ & $0.0048$\\
    \textbf{3} & $100$ & $1000$ & $3$ & $0.8$ & $4$ & $0.1172$ & $0.0072$\\
    \textbf{4} & $150$ & $150$ & $5$ & $0.8$ & $2$ & $0.0320$ & $0.0033$\\
    \textbf{5} & $200$ & $100$ & $6$ & $0.8$ & $2$ & $0.0366$ & $0.0039$\\
  \end{tabular}
}
\end{table}

\begin{figure}[htb]
\centering
\makebox{
\includegraphics[width=0.7\textwidth]{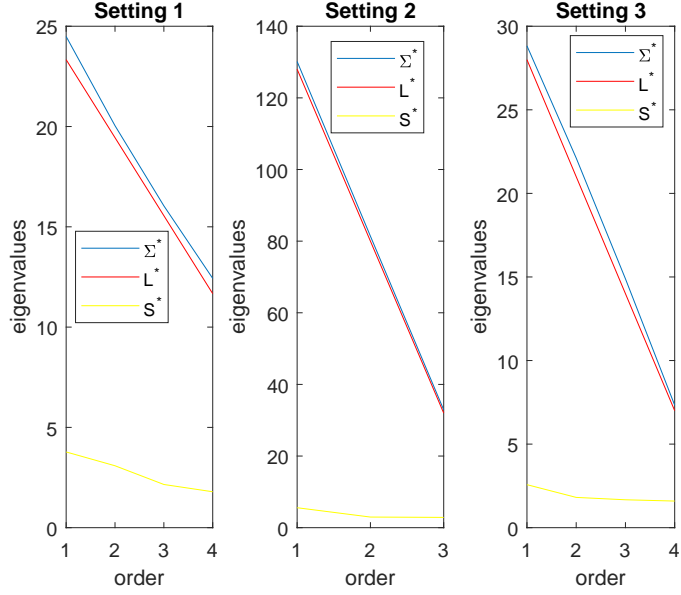}}
\caption{\label{3_1} Eigenvalues of $L$, $S$, $\Sigma$ - Settings 1,2,3}
\end{figure}

\begin{figure}[htb]\label{}
\centering
\makebox{
\includegraphics[width=0.7\textwidth]{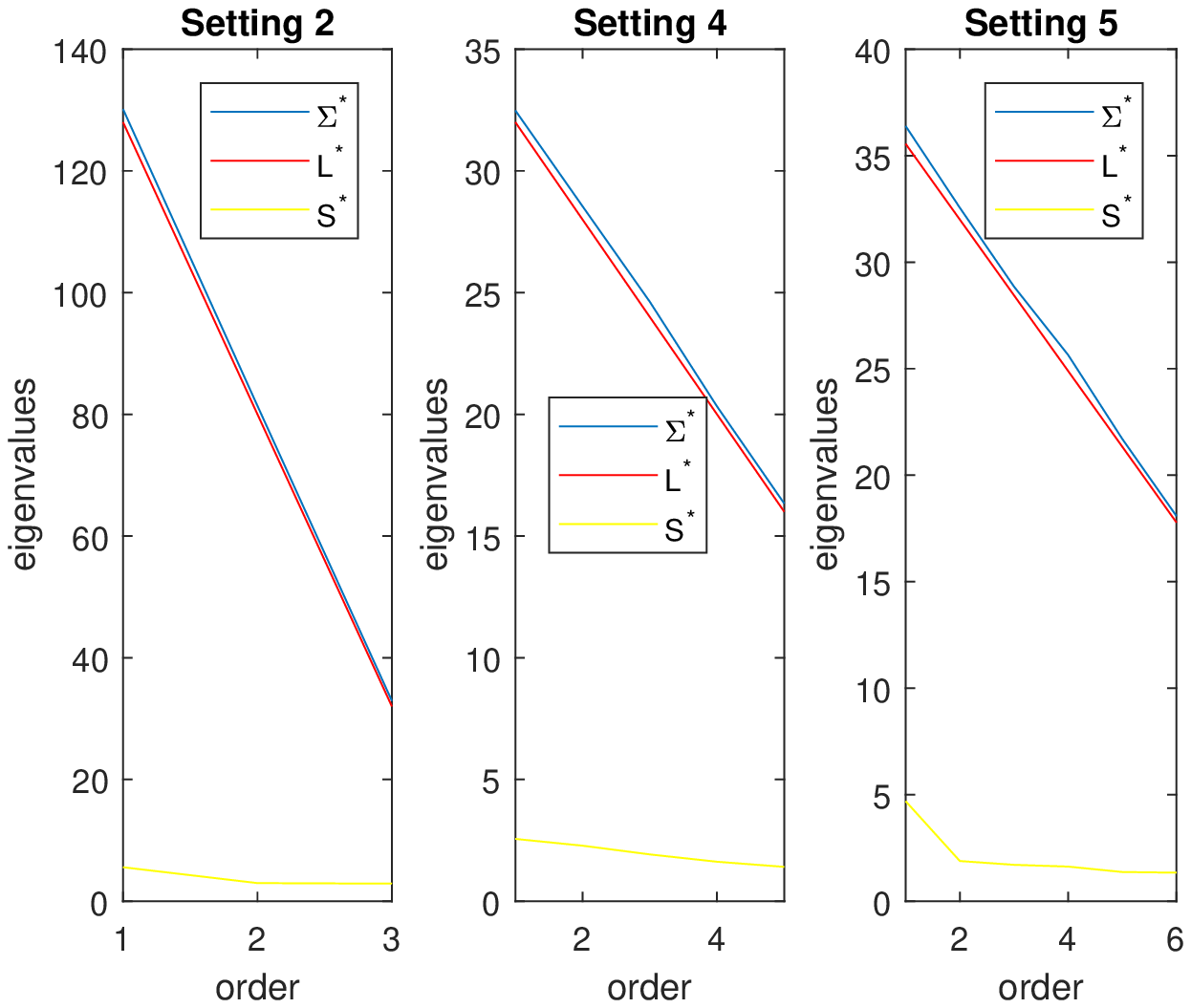}}
\caption{\label{3_2} Eigenvalues of $L$, $S$, $\Sigma$ - Settings 2,4,5}
\end{figure}

In Table \ref{settsumm} we summarize the features of our settings.
Settings 1, 2 and 3 vary according to the degree of spikiness and sparsity.
Setting 4 and 5 are intermediately spiked and sparse, and vary according to the ratio $p/n$.
In particular, Settings 1,2 and 3 have $p/n=0.1$, while Setting 4 has $p/n=1$ and Setting 5 $p/n=2$.
The described features are pointed out in Figures \ref{3_1} and \ref{3_2}, which show the degree of spikiness and sparsity across settings.

\begin{table}
  \caption{\label{specond} Simulated settings: spectral norms and condition numbers}
  \centering
  \fbox{
  \begin{tabular}{ccccccc}
    \textbf{Setting} & \em $||{L}^*||$ & \em $||{S}^*||$ & \em $||{\Sigma}^*||$ & \em $cond_L$ & \em $cond_S$ & \em $cond_{\Sigma}$\\
    \textbf{1} & 23.33 & 3.78 & 24.49 & 2 & 2.26e+07 & 9.49e+07\\
    \textbf{2} &  128 & 5.58 & 130.14 & 4 & 2.53e+05 & 4.07e+06\\
    \textbf{3} & 28 & 2.57 & 28.83 & 4 & 3.80e+07 & 4.04e+07\\
    \textbf{4} & 32 & 2.56 & 32.48 & 2 & 2.35e+13 & 1.58e+10\\
    \textbf{5} & 35.56 & 4.69 & 36.39 & 2 & 1.17e+13 & 3.09e+09\\
  \end{tabular}
}
\end{table}

\begin{table}
  \caption{\label{settsumm} Simulated settings: a summary}
  \centering
  \fbox{
  \begin{tabular}{cccc}
    \textbf{Setting} & \em $p/n$ & \em \mbox{\textbf{spikiness}}& \em \mbox{\textbf{sparsity}}\\
    \textbf{1} &  $0.1$ & \mbox{low} &\mbox{high}\\
    \textbf{2} & $0.1$ & \mbox{middle} & \mbox{middle}\\
    \textbf{3} & $0.1$ & \mbox{high} & \mbox{low}\\
    \textbf{4} & $1$ & \mbox{middle} & \mbox{middle}\\
    \textbf{5} & $2$ & \mbox{middle} & \mbox{middle}\\
  \end{tabular}
}
\end{table}

Our objective (\ref{func:ob1}) is minimized according to an
alternate thresholding algorithm, which is a singular value thresholding (SVT, \cite{cai2010singular})
plus a soft thresholding one \citep{daubechies2004iterative}.
In order to speed convergence, Nesterov's acceleration scheme for composite gradient mapping minimization problems (\cite{nesterov2013gradient}) is applied.
Given a prescribed precision level $\varepsilon$, the algorithm assumes the form \citep{luo2011recovering}:
\begin{enumerate}
\item Set $({L}_{0},{S}_{0})=(diag({\Sigma}_n),diag({\Sigma}_n))/2$, $\eta_0=1$.
\item Initialize ${Y}_{0}={L}_{0}$ and ${Z}_{0}={S}_{0}$. Set $t=1$.
\item \textbf{Repeat}: compute $\frac{\partial \frac{1}{2}||{Y}_{t-1}+{Z}_{t-1}-{\Sigma}_n||^2_{Fro}}{\partial {Y}_{t-1}}=\frac{\partial \frac{1}{2}||{Y}_{t-1}+{Z}_{t-1}-{\Sigma}_n||^2_{Fro}}{\partial {Z}_{t-1}}={Y}_{t-1}+{Z}_{t-1}-{\Sigma}_n$.
\item Apply the SVT operator $T_{\psi}$ to ${Y}_{(t-1)}- \frac{1}{2}({Y}_{t-1}+{Z}_{t-1}-{\Sigma}_n)$ and set ${L}_{t}={U}{D}_\psi {U}'$.
\item Apply the soft-thresholding operator $T_\rho$ to ${M}={Z}_{(t-1)}- \frac{1}{2}({Y}_{t-1}+{Z}_{t-1}-{\Sigma}_n)$ and set ${S}_{t}=T_\rho({M})$.
\item Set $({Y}_{t},{Z}_{t})=({L}_{t},{S}_{t})+\frac{\eta_{t-1}-1}{\eta_{t}}[({L}_{t},{S}_{t})-({L}_{t-1},{S}_{t-1})]$
where $\eta_{t}=\frac{1+\sqrt{1+4 \eta_{t-1}^2}}{2}$.
\item \textbf{Until} the convergence criterion $\frac{||{L}_{t}-{L}_{t-1}||_F}{||1+{L}_{t-1}||_F}+\frac{||{S}_{t}-{S}_{t-1}||_F}{||1+{S}_{t-1}||_F} \leq \varepsilon$.
\end{enumerate}
The reported scheme allows to achieve a convergence speed proportional to $O(t^2)$. We define $t^{*}$ the number of steps needed for convergence.
We set ${Y}_{pre}={Y}_{t^{*}-1}- \frac{1}{2}({Y}_{t^{*}-1}+{Z}_{t^{*}-1}-{\Sigma}_n)$ and ${Z}_{pre}={Z}_{t^{*}-1}- \frac{1}{2}({Y}_{t^{*}-1}+{Z}_{t^{*}-1}-{\Sigma}_n)$.
The computational cost of the solution algorithm is proportional to $\frac{p^4}{\sqrt{\varepsilon}}$, where $\varepsilon$ is the required precision,
while POET has the cost of a full-SVD (proportional to $p^3$).
For more details see \cite{farne2016large}.

Lots of quantities are computed in order to describe comparatively
the performance of the three methods 
on the same data.
We call the low rank estimate $\hat{{L}}$, the sparse estimate $\hat{{S}}$, and
the covariance matrix estimate $\hat{{\Sigma}}=\hat{{S}}+\hat{{L}}$.
The error norms used are:
\begin{enumerate}
\item
$\mbox{\em{Loss}}= ||\hat{{S}}-{S}^*||_{Fro} + ||\hat{{L}}-{L}^*||_{Fro}$,
\item
$\mbox{\em{Total Loss}} = ||\hat{{\Sigma}}-{\Sigma}^*||_{Fro}$,
\item $\mbox{\em{Sample Total Loss}} = ||\hat{{\Sigma}}-{{\Sigma}_n}||_{Fro}.$
\end{enumerate}
The estimated proportion of total variance $\hat{\theta}$ and the residual
covariance proportion $\hat{\rho}_{corr}$ are computed. 
The performance of $\hat{{S}}$ is assessed by the
following measures. Let us denote by $nz$ the number of non-zeros in
$\hat{{S}}$ (recall that $s$ is the number of non-zeros in ${S}^*$),
by $fp$ the false non-zeros, by $fn$ the false zeros, by $fpos$ the false positive
and by $fneg$ the false negative elements.
We define:
\begin{enumerate}
\item the estimated proportion of non-zeros $perc_{nz}=nz/numvar$, where
$numvar=p(p-1)/2$ is the number of off-diagonal elements,
\item the \emph{error} measure: $err={(fp+fn)}/{numvar}$,
\item $errplus={(fpos+fneg)}/{s}$, which is the same as $err$ but computed for non-zeros only,
distinguishing between positive and negative in the usual way.
\item the overall error rate $errtot$ using the number of false zeros, false positive, and false negative elements:
$errtot={(fpos+fneg+fn)}/{numvar}.$
\end{enumerate}

The correct classification rates of (true) non-zeros and zero elements (denoted respectively by $sens$ and $spec$)
are derived, as well as the correct classification rates of positive and negative elements separately considered (denoted respectively by $senspos$ and $specpos$).

\subsection{Simulation results}\label{simres}

\begin{figure}[htb]
\centering
\makebox{
\includegraphics[width=0.7\textwidth]{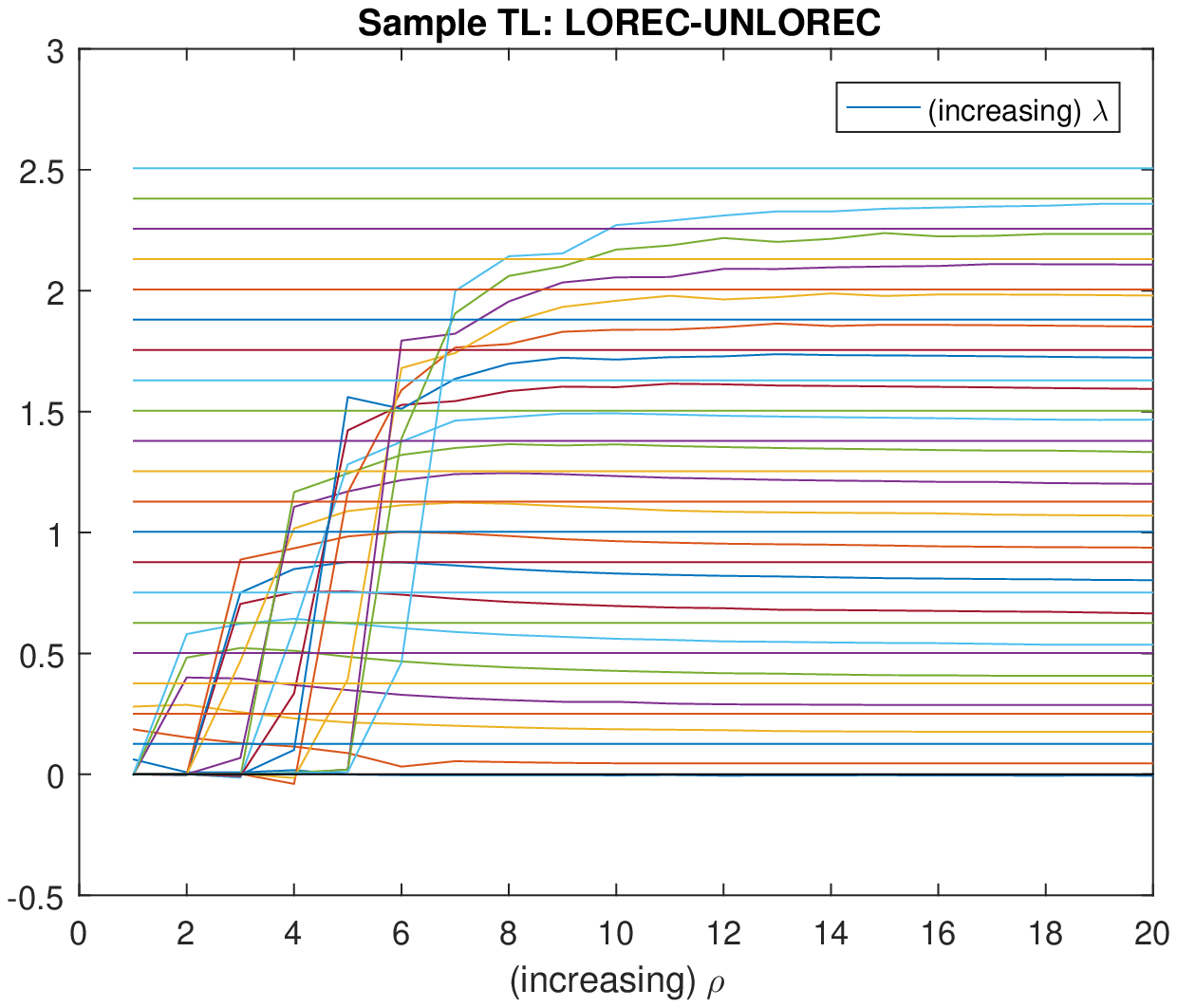}}
\caption{\label{diffTL_s} \emph{Sample Total Loss} difference - $\hat{{\Sigma}}_{LOREC}$ and $\hat{{\Sigma}}_{UNALCE}$ - {Setting 1}}

\end{figure}

\begin{figure}[htb]
\centering
\makebox{
\includegraphics[width=0.7\textwidth]{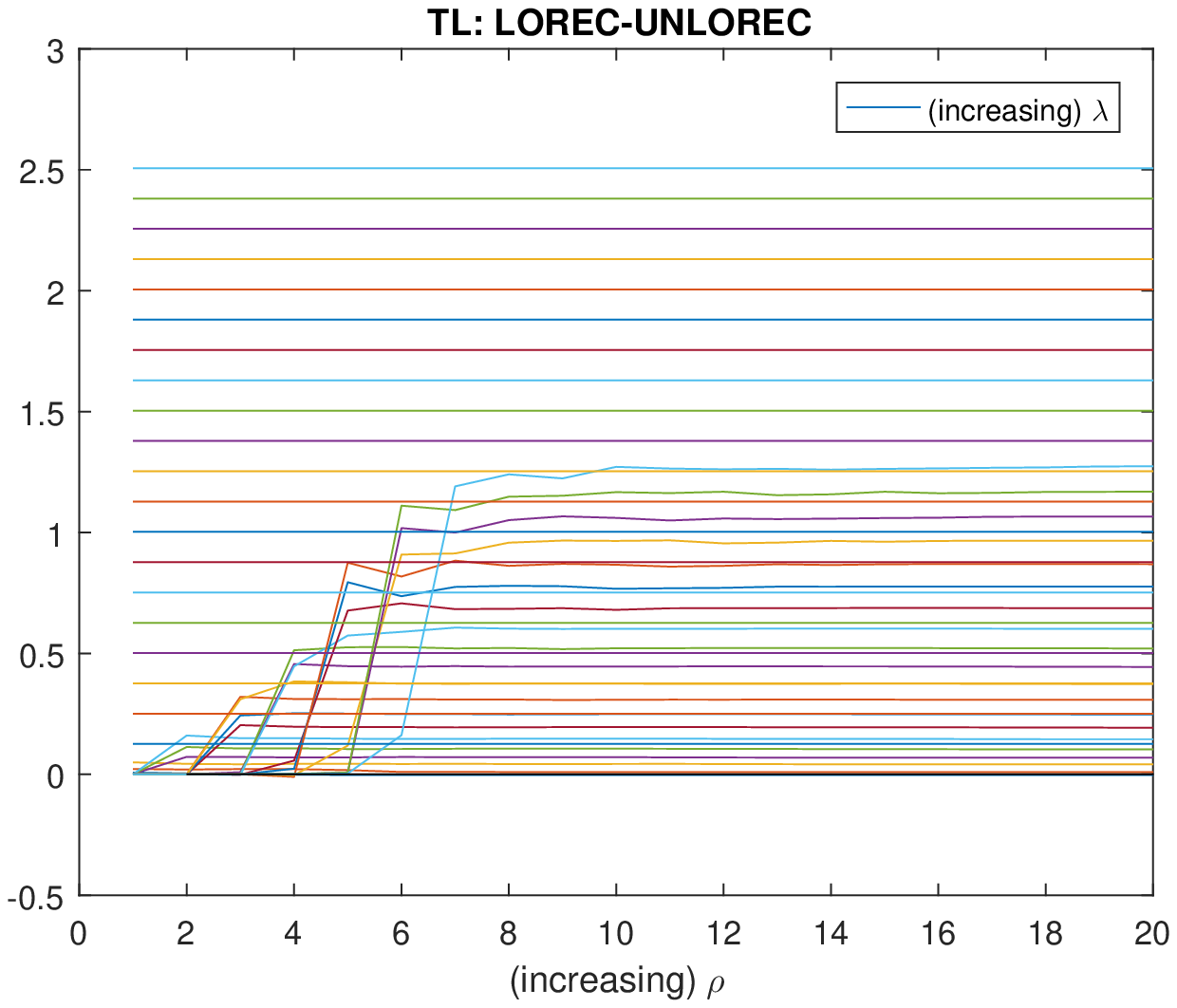}}
\caption{\label{diffTL} \emph{Total Loss} difference - $\hat{{\Sigma}}_{LOREC}$ and $\hat{{\Sigma}}_{UNALCE}$ - {Setting 1}}
\end{figure}

We start analyzing the performance of $\hat{{\Sigma}}_{UNALCE}$ in comparison to the one of $\hat{{\Sigma}}_{LOREC}$
on our reference setting ({Setting 1}).
In Figure \ref{diffTL_s} and \ref{diffTL} we report the differences between the Sample Total Losses and the
Total Losses of LOREC and UNALCE for a grid of $20\times20=400$ threshold pairs.
We note that the gain is positive everywhere, with the exception of the threshold pairs which do not return the exact rank
(because they do not satisfy the range of Theorem  \ref{thmMinetop}).
This pattern is more remarkable for \emph{Sample Total Loss} than for \emph{Total Loss}.
For both losses and each $\psi$, we note that, as explained, the gain across $\rho$ never overcomes its maximum $\sqrt{r}\psi$
(plotted for each $\psi$).

\begin{figure}[htb]
\centering
\makebox{
\includegraphics[width=0.7\textwidth]{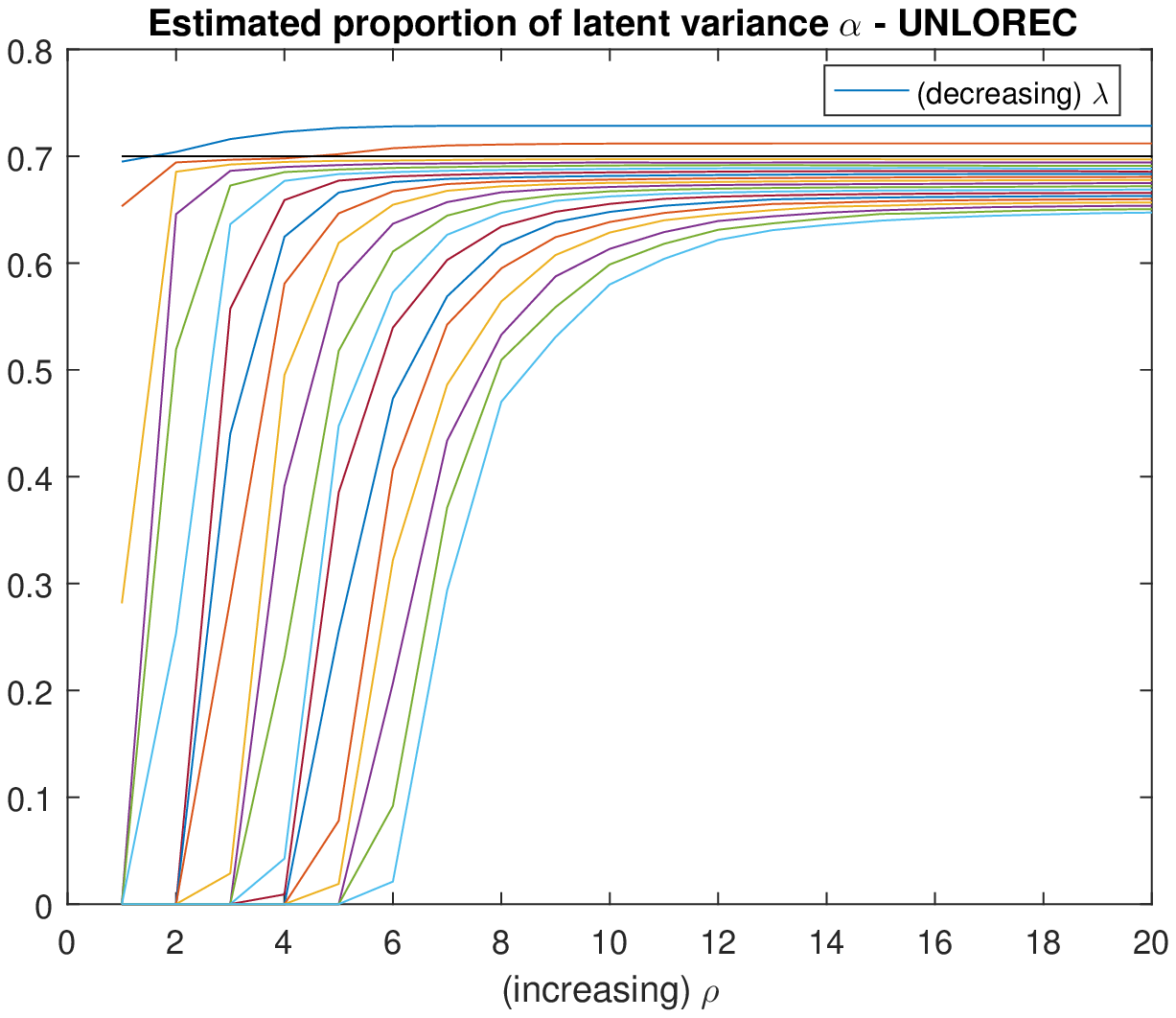}}
\caption{\label{alphahat} Estimated proportion of latent variance - $\hat{{\Sigma}}_{UNALCE}$ - {Setting 1}}
\end{figure}

\begin{figure}[htb]
\centering
\makebox{
\includegraphics[width=0.7\textwidth]{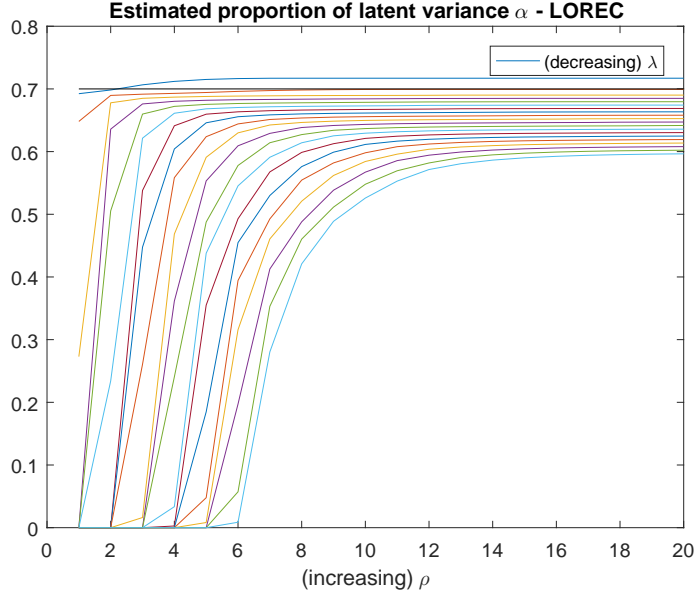}}
\caption{\label{alphahatL} Estimated proportion of latent variance - $\hat{{\Sigma}}_{LOREC}$ - {Setting 1}}
\end{figure}

In Figure \ref{alphahat} we report the plot of the estimated proportion of latent variance $\theta$
across thresholds for $\hat{{\Sigma}}_{UNALCE}$
(in solid line the true $\theta=0.7$).
In Figure \ref{alphahatL} the same plot is reported for $\hat{{\Sigma}}_{LOREC}$.
The shape is exactly the same as for $\hat{{\Sigma}}_{UNALCE}$, the only difference is that all patterns are negatively shifted.
In particular, $\hat{\theta}$ gets closer to $\theta$ for $\hat{{\Sigma}}_{UNALCE}$ respect to $\hat{{\Sigma}}_{LOREC}$
in correspondence to all threshold combinations.

\begin{table}
        \caption{\label{loss_1} Simulation results over 100 runs}
        \centering
        \fbox{
        \begin{tabular}{ccccccc}
             & \multicolumn{2}{c}{Setting 1} & \multicolumn{2}{c}{Setting 2} & \multicolumn{2}{c}{Setting 3}\\
             & UNALCE & POET & UNALCE & POET & UNALCE & POET\\
            $\hat{\alpha}$ & \textbf{0.6952} & 0.7314 & \textbf{0.6955} & 0.7324 & \textbf{0.7987} & 0.8151\\
            $\hat{\rho}_{\hat{S}}$  & \textbf{0.0034} & 0.0000  & \textbf{0.0071} & 0.0000 & \textbf{0.0035} & 0.0000 \\
            $prop_{nz}$ & \textbf{0.0299} & 0.0003 & \textbf{0.0915} & 0.0079 & \textbf{0.1287} & 0.0132\\
            $TL$ & \textbf{6.98} & 7.39 & \textbf{11.69} & 11.70 & \textbf{9.95} & 10.47\\
            $SampleTL$ & \textbf{0.72} & 2.79  & \textbf{0.57} & 2.22 & \textbf{1.26} & 3.85\\
            $Loss$ & \textbf{7.63} & 9.32 & \textbf{12.36} & 12.95 & \textbf{11.39} & 13.26 \\
            $Loss_L$ & \textbf{6.91} & 7.58 & \textbf{11.59} & 11.62 & \textbf{9.85} & 10.74 \\
            $Loss_S$ & \textbf{0.72} & 1.74  & \textbf{0.78} & 1.33 & \textbf{1.55} & 2.52\\
        \end{tabular}
        }
\end{table}

\begin{table}
        \caption{\label{loss_2} Simulation results over 100 runs}
        \centering
        \fbox{
        \begin{tabular}{ccccccc}
             & \multicolumn{2}{c}{Setting 2} & \multicolumn{2}{c}{Setting 4} & \multicolumn{2}{c}{Setting 5}\\
             & UNALCE & POET & UNALCE & POET & UNALCE & POET\\
            $\hat{\alpha}$ & \textbf{0.6955} & 0.7324  & \textbf{0.7980} & 0.8233 & \textbf{0.7932} & 0.8284\\
            $\hat{\rho}_{\hat{S}}$   & \textbf{0.0071} & 0.0000 & \textbf{0.0022} & 0.0000 & \textbf{0.0004} & 0.0000\\
            $prop_{nz}$ & \textbf{0.0915} & 0.0079  & \textbf{0.0164} & 0.0115 & \textbf{0.0015} & 0.0010\\
            $TL$  &  \textbf{11.69} & 11.70 & \textbf{13.02} & 13.31 & \textbf{20.92} & 21.41\\
            $SampleTL$  & \textbf{0.57} & 2.22 & \textbf{1.94} & 2.90 & \textbf{3.91} & 4.38\\
            $Loss$ & \textbf{12.36} & 12.95 & \textbf{14.25} & 14.89 & \textbf{22.49} & 23.97\\
            $Loss_L$ & \textbf{11.59} & 11.62 & \textbf{12.93} & 13.38 & \textbf{20.85} & 21.53\\
            $Loss_S$ & \textbf{0.78} & 1.33  & \textbf{1.32} & 1.51 & \textbf{1.65} & 2.44\\
        \end{tabular}
        }
    \end{table}

    \begin{table}
        \caption{\label{err_1} Simulation results over 100 runs}
        \centering
        \fbox{
        \begin{tabular}{ccccccc}
             & \multicolumn{2}{c}{Setting 1} & \multicolumn{2}{c}{Setting 2} & \multicolumn{2}{c}{Setting 3}\\
             & UNALCE & POET & UNALCE & POET & UNALCE & POET\\
$err$ &\textbf{0.0195}&	0.0242&	\textbf{0.0967}&	0.1250&	\textbf{0.0626}&	0.0808\\
$errplus$ &0.111&	\textbf{0.0000}&	0.0183&	\textbf{0.0001}&	0.0193&	\textbf{0.0003}\\
$errtot$ &\textbf{0.0093}&	0.0238&	\textbf{0.0507}&	0.1172&	\textbf{0.0270}&	0.0676\\
$senspos$ &\textbf{0.7019}&	0.0000&	\textbf{0.6977}&	0.0002&	\textbf{0.6077}&	0.0010\\
$specpos$ &\textbf{0.7105}&	0.0000&	\textbf{0.6909}&	0.0000&	\textbf{0.6294}&	0.0000\\
$spec$ &0.9869&	\textbf{0.9997}&	0.9536&	\textbf{0.9911}&	0.9387&	\textbf{0.9859}\\
        \end{tabular}
    }
    \end{table}

\begin{table}
        \caption{\label{err_2} Simulation results over 100 runs}
        \centering
        \fbox{
        \begin{tabular}{ccccccc}
             & \multicolumn{2}{c}{Setting 2} & \multicolumn{2}{c}{Setting 4} & \multicolumn{2}{c}{Setting 5}\\
                         & UNALCE & POET & UNALCE & POET &  UNALCE & POET\\
$err$ &\textbf{0.0967}&	0.1250&	\textbf{0.0321}&	0.0435&	\textbf{0.0359}&	0.0375\\
$errplus$ &	0.0183&	\textbf{0.0001}&	0.0147&	\textbf{0.0001}&	0.0025&	\textbf{0.0000}\\
$errtot$ &\textbf{0.0507}&	0.1172&	\textbf{0.0245}&	0.0320&	\textbf{0.0356}&	0.0366\\
$senspos$ &\textbf{0.6977}&	0.0002&	\textbf{0.2318}&	0.0001&	\textbf{0.0284}&	0.0000\\
$specpos$ &	\textbf{0.6909}&	0.0000&	\textbf{0.2414}&	0.0000&	\textbf{0.0262}&	0.0000\\
$spec$ &0.9536&	\textbf{0.9911}&	\textbf{0.9915}&	0.9882&	\textbf{0.9995}&	0.9990\\
        \end{tabular}
        }
    \end{table}

\begin{table}
        \caption{\label{cond_1} Simulation results over 100 runs}
        \centering
        \fbox{
        \begin{tabular}{ccccccc}
             & \multicolumn{2}{c}{Setting 1} & \multicolumn{2}{c}{Setting 2} & \multicolumn{2}{c}{Setting 3}\\
             & UNALCE & POET & UNALCE & POET &  UNALCE & POET\\
$eig_{\hat{\Sigma}}$ & \textbf{5.50} & 5.74 & \textbf{11.14} & 11.62 & \textbf{5.64}	& 6.07\\
$eig_{\hat{S}}$  &\textbf{0.29}	&1.55	&	\textbf{0.26}	&1.17 &\textbf{0.43}&	1.86\\
$eig_{\hat{L}}$ & 7.75	&\textbf{7.16}	&	\textbf{14.52}	&15.24 &\textbf{5.65}&	6.16\\
$cond_{\hat{\Sigma}}$ &\textbf{104110}	& 34048 &\textbf{114210}	&14452	&\textbf{2207400}	&1141400\\
$cond_{\hat{S}}$ &\textbf{21571}	&4776.3&\textbf{65628}	&1310.2 &\textbf{125170}	&40407	\\
$cond_{\hat{L}}$ &1.32	&\textbf{1.32}	&1.54	&\textbf{1.55} &4.07	&\textbf{3.97}\\
$||\hat{\Sigma}||$ &20.84&	\textbf{21.84}	&21.90	&\textbf{22.59} &\textbf{130.22}	&131.58	\\
$|\hat{|S}||$ &\textbf{3.77}	&2.75	&\textbf{2.66}	&1.90 &\textbf{5.67}	&4.15	\\
$|\hat{|L}||$ &19.84	&\textbf{21.00}	&21.00	&\textbf{22.01} &\textbf{128.50}	&130.39	\\
        \end{tabular}
        }
    \end{table}


\begin{table}
        \caption{\label{cond_2} Simulation results over 100 runs}
        \centering
        \fbox{
        \begin{tabular}{ccccccc}
             & \multicolumn{2}{c}{Setting 2} & \multicolumn{2}{c}{Setting 4} & \multicolumn{2}{c}{Setting 5}\\
                         & UNALCE & POET & UNALCE & POET &  UNALCE & POET\\
$eig_{\hat{\Sigma}}$ & \textbf{11.14} & 11.62 &	\textbf{6.06} &	6.24 &	\textbf{10.06}	 &10.57\\
$eig_{\hat{S}}$ &	\textbf{0.26}	&1.17 &	\textbf{0.49} &	1.15 &	\textbf{0.81}	 &1.92\\
$eig_{\hat{L}}$ &	\textbf{14.52}	&15.24 &	\textbf{6.07} &	6.34 &	\textbf{10.43}	 &10.57\\
$cond_{\hat{\Sigma}}$ &\textbf{114210}	&14452 &	\textbf{28321} &	192790 &	12857&	\textbf{20171}\\
$cond_{\hat{S}}$  &\textbf{65628}	&1310.2&	\textbf{2469} &	1132.2	 &1406.9&	\textbf{1430}\\
$cond_{\hat{L}}$ &1.54	&\textbf{1.55} &	2.41 &	\textbf{2.35}  &2.99	 &\textbf{2.85}\\
$||\hat{\Sigma}||$  &21.90	&\textbf{22.59} &	\textbf{35.34} &	36.03 &	\textbf{42.48} &	43.57\\
$||\hat{S}||$ &\textbf{2.66}	&1.90	 &\textbf{2.73}	&1.90&	\textbf{4.49} &	3.15\\
$||\hat{L}||$ &21.00	&\textbf{22.01}	&\textbf{35.68}	& 34.88 &	\textbf{43.17}	 &42.00\\
        \end{tabular}
        }
    \end{table}

Table \ref{loss_1} contains some results about fitting measures across different degrees of spikiness (Settings 1,2,3).
It is clear that UNALCE outperforms POET concerning all losses,
and shows the superior performance of UNALCE concerning the proportion of latent variance, of residual covariance, of detected non-zeros.
The same pattern can be deduced from Table \ref{loss_2}, which contains the same results across different ratios $p/n$ (Settings 2,4,5).
Nevertheless, we note that the gap progressively decreases as $p/n$ increases, due to the increased consistency with POET assumptions.
We note, for instance, that the proportion of residual covariance is underestimated also by UNALCE for $p/n=2$.
At the same time, the performance of the proportion of residual variance detected by POET is upper biased,
due to the natural bias of sample eigenvalues, and the bias decreases as the degree of spikiness increases.

Tables \ref{err_1} and \ref{err_2} contain the error measures about the detection of the residual pattern across the degree of spikiness and the ratio $p/n$ respectively.
We note that POET, due to the lack of algebraic consistency, is completely unable to classify positive and negative elements. On the contrary, UNALCE shows a recovery rate around $70\%$ when $p/n$ is small, while the detection capability deteriorates as $p/n$ increases.

Tables \ref{cond_1} and \ref{cond_2} report the Euclidean distance between the vectors of estimated and true eigenvalues (denoted by $eig$),
the condition number of the estimates and the estimated spectral norms.
This Table can be compared to Table \ref{specond} which contains the true spectral norms and condition numbers across settings.
All statistics are generally in favour of UNALCE with some notable exceptions motivated by our theory.
If $p/n$ is low and the eigenvalues are not spiked, the spectral norm tend to be underestimated by UNALCE, because the eigenvalues tend to be smaller and more concentrated.
On the contrary, UNALCE may overestimate the condition number of ${L}^*$ and ${\Sigma}^*$
if $p/n$ is large, because the guarantee required for positive definiteness is stronger.

To sum up, our UNALCE estimator outperforms POET concerning fitting and conditioning properties, detection of sparsity pattern,
and eigen-structure recovery.
We note that POET does not detect positive and negative elements at all.
This is because it only has parametric consistency and not also the algebraic one.
In addition, in order to obtain a positive definite estimate, cross validation selects a very high threshold for POET, and this causes the sparse estimate to be almost completely diagonal if $p/n$ is large.
On the contrary, the mathematical optimization procedure of UNALCE gets closer to the target and ensures to catch the algebraic spaces behind the two components.








\section{Proofs}\label{proofs}

\subsection{Proof of Theorem \ref{thmMinetop}}

Theorem \ref{thmMinetop} relies on proving the equivalent of the following Lemma by \cite{fan2013large}
\begin{Lemma}
\begin{equation}||{{{\Sigma}}}_{n}-{\Sigma}^{*}||\leq C \left(\frac{p}{\sqrt{n}}\right)\end{equation} \label{plusmore2}
\end{Lemma}
under our assumption setting.
Lemma (\ref{plusmore2}) in turn relies on the following Lemmas by \cite{fan2011high}:
\begin{Lemma}
$$\max_{i,j \leq r} \biggl\vert  \frac{1}{T} \sum_{t=1}^T f_{it}f_{jt}-E(f_{it}f_{jt})\biggl\vert \leq C \left(\frac{1}{\sqrt{n}}\right)$$ \label{Lemma4_1}
\end{Lemma}
\begin{Lemma}
$$\max_{i,j \leq r} \biggl\vert  \frac{1}{T} \sum_{t=1}^T s_{it}s_{jt}-E(s_{it}s_{jt})\biggl\vert \leq C \left(\frac{\log (p)}{\sqrt{n}}\right)$$ \label{Lemma4_2}
\end{Lemma}
\begin{Lemma}
$$\max_{i,j \leq r} \biggl\vert \frac{1}{T} \sum_{t=1}^T f_{it}s_{jt}\biggl\vert \leq C \left(\frac{\log (p)}{\sqrt{n}}\right).$$ \label{Lemma4_3}
\end{Lemma}
As a consequence, we need to explore how Lemmas \ref{Lemma4_1}, \ref{Lemma4_2}, \ref{Lemma4_3} change under our assumptions. 

The first step of the proof consists in decomposing ${E}_n={\Sigma}_n-{\Sigma}^*$ in its four components:
$${E}_n={\Sigma}_n-{\Sigma}^*={D}_1+{D}_2+{D}_3+{D}_4$$ where:
$${D}_1=\left(n^{-1} {B} \sum_{i=1}^n {f}_i {f}_i'-{I}_r\right){B}'$$
$${D}_2=n^{-1} \left(\sum_{i=1}^n {\epsilon}_i {\epsilon}_i'-{S}^*\right)$$
$${D}_3={B} n^{-1}\sum_{i=1}^n {f}_i {\epsilon}_i'$$
$${D}_4={D}_3',$$
where ${{f}}_i$ and ${{\epsilon}}_i$ are respectively the vectors of factor scores and residuals for each observation.

From the inequality $||{B}'{\Sigma}^{*-1}{B}|| \leq |cov ({f}) ^{-1}|$ (\cite{fan2008high}, page 194, Assumption (B)), Lemma \ref{Lemma4_1} follows under Assumption \ref{tails}. Therefore, Lemma \ref{Lemma4_1} is unaffected. 
Consequently, following \cite{fan2013large}, we can argue that
$$||{D}_1|| \leq C {r}\sqrt{\frac{\log(r)}{n}}||{BB'}||\leq C \left(p^{\alpha}\sqrt{\frac{1}{n}}\right)$$
because $r\leq \delta_3 \log{p^{3\delta}}$ (Assumption \ref{pr}) and $||{BB'}||=O(p^{\alpha})$ (Assumption \ref{spikyPOETnew}).

In order to show how Lemma \ref{Lemma4_2} changes, we need to recall some key results of \cite{bickel2008covariance}.
Differently from Luo's approach, in that setting (as in ours and in the POET one) the sparsity assumption is imposed to ${S}^{*}$, and not to ${\Sigma}^{*}$.
While in \cite{fan2013large} the assumption $\max_{i\leq p} \sum_{j \leq p} |s^*_{ij}|^q = o(p)$, $q\in[0,1]$ is needed
to ensure POET consistency,
here Assumption \ref{tails2} prescribes $\max \sum_{j \leq p} \mathbbm{1}(s^*_{ij}=0)\leq \delta_2 p^{\delta}$.

Consider now the uniformity class of sparse matrices in \cite{bickel2008covariance} (with $q=0$):
\begin{equation}
\left\{{S}^{*}: s^{*}_{ii} \leq c_3, \: \max_i \sum_{j}  \mathbbm{1}(s^{*}_{ij} \ne 0)  \leq c_0(p), \: \forall i\right\}.\label{def:S_eps2}
\end{equation}
Under Assumption \ref{tails2} this class is no longer appropriate, because
we can no longer write (see \cite{bickel2008covariance}, page 2580)
$$\lambda_{max}({S}^{*})\leq \max_i \sum_{j} \mathbbm{1}(s^{*}_{ij} \ne 0) \leq c_3 c_0(p),$$
since the quantity $c_0(p)$ 
no longer scales to $p$ but to $p^{\delta}$. 
Therefore, we need to replace $c_0(p)$ by $C(p^{\delta-1})$ in the proof which derives the rate of the sample covariance matrix $\hat{{S}}_n$
under class (\ref{def:S_eps2})
(see \cite{bickel2008covariance}, page 2582),
thus proving under Assumptions \ref{tails} and \ref{tails2} that: 
\begin{Lemma}\label{probinf2}
$$||{\hat{{S}}}_n-{S}^*||_{\infty}\leq C\left(p^{\delta-1}\sqrt{\frac{\log{p}}{n}}\right).$$
\end{Lemma}

Using Lemma \ref{probinf2}, we can derive 
$$||{D}_2|| \leq Cp(p^{\delta-1}) \left(\sqrt{\frac{\log(p)}{n}}\right)= C \left(p^{\delta}\sqrt{\frac{\log{p}}{n}}\right),$$
because $||{D}_2||\leq p ||{D}_2||_{\infty}.$
Since $log(p) \ll n$, we can write
\begin{equation}||{D}_2|| \leq C p||{\hat{{S}}}_n-{S}^*||_{\infty} \sqrt{\frac{\log{p}}{n}}=C p (p^{\delta-1})
\sqrt{\frac{\log{p}}{n}}= C \left(p^{\delta} \frac{1}{\sqrt{n}}\right).\end{equation}

To conclude, we analyze Lemma \ref{Lemma4_3}:
$$\max_{i\leq r,j\leq p}  \biggl\vert\frac{1}{n} \sum_{k=1}^n f_{ik}s_{jk} \biggl\vert
\leq \frac{1}{\sqrt{n}}\sum_{k=1}^n \max_i{|f_{ik}|}\frac{1}{\sqrt{n}}\max_{j}\sum_{k=1}^n |s_{jk}| \leq \sqrt{\frac{r}{n}} p p^{\delta-1} \sqrt{\frac{\log{p}}{n}},$$

Exploiting Assumption \ref{pr} 
we obtain
$\sqrt{\frac{r}{n}}\leq C(p^{-3\delta})$.
Therefore, the bound above becomes $C\left(p^{-2\delta}\sqrt{\frac{\log{p}}{n}}\right)$.

Applying the recalled proof strategy to ${D}_3$ we obtain 
$$||{D}_3|| \leq \biggl\vert \frac{1}{n} \sum_{i=1}^n {f}_i{u}_i'\biggl\vert \times ||{B}|| \leq C\left(p^{-2\delta}\sqrt{\frac{\log{p}}{n}}\right)\left(p^{\frac{\alpha}{2}}\right)=C\left(p^{\frac{\alpha}{2}-2\delta}\sqrt{\frac{\log{p}}{n}}\right),$$
because $||{B}||=O(p^{\alpha/{2}})$ by Assumption \ref{spikyPOETnew}. The condition $log(p)\ll n$ finally leads to:
\begin{equation}||{D}_3||\leq C \left(\frac{p^{\frac{\alpha}{2}-2\delta}}{\sqrt{n}}\right).\label{toptop}\end{equation}

Therefore, the following bound is proved
\begin{equation}||{{{\Sigma}}}_{n}-{\Sigma}^{*}||\leq C\left(\frac{p^{\alpha}}{\sqrt{n}}\right),\label{ratetop}\end{equation}
because $\delta \leq \alpha$ from Assumption \ref{alg}. In fact, if $\delta > \alpha$ the condition of Theorem \ref{thmMinetop} $\lambda_r({L}^{*})> C_2 \frac{\psi}{\xi^2(T)}$ would result in $\lambda_r({L}^{*})>C_2 p^{\delta}$, thus violating Assumption \ref{spikyPOETnew}.





In other words, the bound (\ref{ratetop}) means
\begin{equation}P\left(||{E}_n|| \geq C_1 \frac{p^{\alpha}} {\sqrt{n}}\right) \leq 1- C_2 \exp{(-C_3 p^{2\alpha})}.\end{equation}
The proof relies on the combined use of proof tools by \cite{fan2013large}, \cite{fan2011high}, \cite{fan2008high} and \cite{bickel2008covariance}.

Exploiting the basic property $||.||_{\infty}\leq ||.||_2$
and the minimum for $\gamma$ in the range of Theorem \ref{thmMinetop}, we can simply write
\begin{equation}P\left(||{E}_n||_{\infty} \geq C_1 \xi(T) \frac{p^{\alpha}} {\sqrt{n}}\right)\leq 1- C_2 \exp{(-C_3 p^{2\alpha})}.\label{plusfinal}\end{equation}

According to \cite{chandrasekaran2012} and \cite{luo2011high}, the only probabilistic component of the error norm $g_\gamma(\hat{{S}}-{S}^{*},\hat{{L}}-{L}^{*})$ is $g_\gamma({E}_n)$.
Therefore, following the proof of \cite{luo2011high} and setting $\psi=\left(\frac{1}{\xi(T)}\frac{p^{\alpha}}{\sqrt{n}}\right)$,
we can finally prove, under all the assumptions and conditions of Theorem \ref{thmMinetop}, the thesis
\begin{equation}g_\gamma(\hat{{S}}-{S}^{*},\hat{{L}}-{L}^{*})\leq C\frac{1}{\xi(T)}{\frac{p^{\alpha}}{\sqrt{n}}}.\label{gplus}\end{equation}

\subsection{Proof of Corollary \ref{asy}}


The proof directly descends by bound (\ref{ratetop}), because $\frac{p^{\alpha}}{\sqrt{n}}\rightarrow 0$ if and only if $\frac{p^{2\alpha}}{n}=o(1)$ as $\min(p,n) \rightarrow \infty$. As expected, the absolute bound vanishes only in the small dimensional case ($n\gg p^{\alpha}\log(p)$).

\subsection{Proof of Corollary \ref{rc}}
Defined ${\Sigma}_r$ as the covariance matrix formed by the first $r$ principal components,
we know by dual Lidskii inequality that
\begin{eqnarray}
\lambda_r(\hat{{L}}_{UNALCE})={\lambda}_r(\hat{{L}}_{UNALCE}-{\Sigma}_r+{\Sigma}_r)\geq \nonumber\\
\geq{\lambda}_{p-r+1}(\hat{{L}}_{UNALCE}-{\Sigma}_r)+\lambda_r({\Sigma}_r).\nonumber
\end{eqnarray}
We start studying the behaviour of $\lambda_r({\Sigma}_n)$.
From bound (\ref{ratetop}) it descends the following Lemma
\begin{Lemma}\label{samplego}
Let $\hat{\lambda}_r$ be the $r-$th largest eigenvalue of ${\Sigma}_n$. If $\alpha \geq \delta$,
then $\hat{\lambda}_r>C_1 \frac{p^{\alpha}}{\sqrt{n}}$ with probability approaching $1$ for some $C_1>0$.
\end{Lemma}

By Lidskii dual inequality, in fact, we note
$$\lambda_r({\Sigma}_n)=\lambda_r({\Sigma}_n-{\Sigma}^*+{\Sigma}^*)
\geq {\lambda}_p({\Sigma}_n-{\Sigma}^*)+\lambda_r({\Sigma}^*).$$

Applying Lidskii dual inequality to $\lambda_r({\Sigma}^*)$ we have
$\lambda_r({\Sigma}^*)=\lambda_r({L}^*+{S}^*)\geq \lambda_r({L}^*)+ \lambda_p({S}^*)$.
If $\alpha\geq\delta$, we obtain $\lambda_r({\Sigma}^*)\geq \delta_{\alpha} {p^{\alpha}}$ by Assumption \ref{spikyPOETnew}, which means that $\lambda_r({\Sigma}^*)$ is bounded away from $0$ and $\infty$ as we divide by $p^{\alpha}$ for $p\rightarrow \infty$.
Otherwise, we obtain $\lambda_r({\Sigma}^*)\geq 0$.

Applying Weyl's inequality to $({\Sigma}_n-{\Sigma}^*)$ we obtain
${\lambda}_p({\Sigma}_n-{\Sigma}^*)\leq ||{{{\Sigma}}}_{n}-{\Sigma}^{*}||\leq C\left(\frac{p^{\alpha}}{\sqrt{n}}\right)$ by bound (\ref{ratetop}).
Therefore, dividing by $p^{\alpha}$, ${\lambda}_p({\Sigma}_n-{\Sigma}^*)\rightarrow 0$.

Finally, assuming $\alpha\geq\delta$, ${\lambda}_{p-r+1}(\hat{{L}}_{UNALCE}-{\Sigma}_r)$ vanishes asymptotically dividing by $p^{\alpha}$ because both the relative errors of $\hat{{L}}_{UNALCE}$ and $\hat{{L}}_{POET}={\Sigma}_r$ vanish (once assumed that $r$ is known a priori or consistently estimated by UNALCE).

Then Corollary \ref{rc} is proved because $\delta_{\alpha} > 0$.

\subsection{Proof of Theorem \ref{mine}}

Conditioning on ${Y}_{pre}$, ${Z}_{pre}$ and ${\Sigma}_{pre}={Y}_{pre}+{Z}_{pre}$,
we aim to solve $$\min_{{L} \in \hat{\mathcal{B}}(\hat{r}),{S }\in \hat{\mathcal{A}}(\hat{s}),{\Sigma}={L}+{S}} ||{\Sigma}-{\Sigma}_n||^2_{Fro}=||{\Sigma}-{\Sigma}_{pre}+{\Sigma}_{pre}-{\Sigma}_{n}||^2_{Fro}.$$
By Cauchy-Schwartz inequality, it can be shown that \begin{eqnarray}
||{\Sigma}-{\Sigma}_{pre}+{\Sigma}_{pre}-{\Sigma}_{n}||^2_{Fro}\leq \nonumber\\
\leq ||{\Sigma}-{\Sigma}_{{pre}}||^2_{Fro}+
||{\Sigma}_{pre}-{\Sigma}_n||^2_{Fro}.\nonumber \end{eqnarray}
${\Sigma}_{pre}$ solves the problem \begin{equation}\min_{{L} \in \hat{\mathcal{B}}(\hat{r}),{S }\in \hat{\mathcal{A}}(\hat{s}),{\Sigma}={L}+{S}} ||{\Sigma}_{pre}-{\Sigma}_n||^2_{Fro}\nonumber\end{equation} conditioning on the fact that
$\breve{\psi} ||{L}||_{*}+ \breve{\rho}||{S}||_1$ is minimum over the same set.

Then we can write $$||{\Sigma}-{\Sigma}_{{pre}}||^2_{Fro}=||{L}+{S}-{Y}_{pre}-{Z}_{pre}||^2_{Fro}.$$
By Cauchy-Schwartz inequality, it can be shown that \begin{equation}
||{L}+{S}-{Y}_{pre}+{Z}_{pre}||^2_{Fro}\leq ||{L}-{Y}_{pre}||^2_{Fro}+||{S}-{Z}_{pre}||^2_{Fro}.\nonumber
\end{equation}

Hence
\begin{equation}\min_{{L} \in \hat{\mathcal{B}}(\hat{r}),{S}\in \hat{\mathcal{A}}(\hat{s}),{\Sigma}={L}+{S}}  ||{L}+{S}-{Y}_{pre}+{Z}_{pre}||^2_{Fro} \leq \end{equation}
$$\leq \min_{{L} \in \hat{\mathcal{B}}(\hat{r})} ||{L}-{Y}_{pre}||^2_{Fro}+
\min_{{S} \in \hat{\mathcal{A}}(\hat{s})} ||{S}-{Z}_{pre}||^2_{Fro}.$$

The problem in ${L}$ is solved taking out the first $\hat{r}$ principal components of ${Y}_{pre}$. By construction, the solution is $\hat{ U}_{ALCE}(\hat{ D}_{ALCE}+\breve{\psi} {I}_r)\hat{ U}_{ALCE}'=\hat{{L}}_{UNALCE}$.
The problem in ${S}$, assuming that the diagonal of $\hat{ {\Sigma}}_{ALCE}$ is given and the off-diagonal elements of $\hat{{S}}$ are invariant,
leads to: 
\begin{eqnarray}
\min_{{S}\in
\hat{\mathcal{A}}(\hat{s})}||{S}- {Z}_{pre}||^2_{Fro}=\nonumber
\\=\min_{{L} \in \hat{\mathcal{ B}}(\hat{r})}||(\hat{ {\Sigma}}-{L})-( {\Sigma}_{pre}- {Y}_{pre})||^2_{Fro}=\nonumber\\
=\min_{{L} \in \hat{\mathcal{ B}}(\hat{r})}||(\hat{ {\Sigma}}- {\Sigma}_{pre})-({L}- {Y}_{pre})||^2_{Fro}\leq \nonumber\\
||(\hat{ {\Sigma}}- {\Sigma}_{pre})||^2_{Fro}+||({L}- {Y}_{pre})||^2_{Fro}
= {B'}+ {B''}.\nonumber
\end{eqnarray}
The question now becomes: which diagonal elements of ${L}$ ensure the minimum of $ {B'}+ {B''}$?
Term $ {B'}$ is fixed respect to ${L}$, because we are assuming the invariance of diagonal elements in
$\hat{ {\Sigma}}$ ($diag(\hat{ {\Sigma}}_{UNALCE})=diag(\hat{ {\Sigma}}_{ALCE})$).
The minimization of term $ {B''}$, given that $rank({L})=\hat{r}$, falls back into the previous case, i.e. $ {B''}$ is minimum if and only if $\hat{{L}}=\hat{{L}}_{UNALCE}~=~\hat{{U}}_{UNALCE}(\hat{{D}}_{UNALCE}+\breve{\psi}  {I}_r) \hat{{U}}_{UNALCE}'$.

Optimality holds over the cartesian product of the set of all positive semi-definite matrices with rank smaller or equal to $r$,
$\hat{\mathcal{B}}(\hat{r})$,
and the set of all sparse matrices with the same sparsity pattern as $\hat{{S}}_{ALCE}$ such that $diag({S})=diag(\hat{{\Sigma}}_{ALCE}-{L})$, ${L} \in \hat{\mathcal{B}}(\hat{r})$
(we call this set $\hat{\mathcal{A}}_{diag}(\hat{s}))$.


Consequently, we can write:
\begin{eqnarray}
\hat{{S}}_{UNALCE,ii}=\hat{ {\Sigma}}_{ii}-\hat{{L}}_{UNALCE,ii},\: \forall i. \nonumber \\
\hat{{S}}_{UNALCE,ij}=\hat{{S}}_{ij}, \: \forall i \ne j.\nonumber
\end{eqnarray}

\subsection{Proof of Corollary \ref{mine2}}

We know that $||\hat{{L}}_{UNALCE}-\hat{{L}}_{ALCE}||_{2} =\breve{\psi}$.
We can prove that  $$\hat{{L}}_{UNALCE}=\min_{{L} ~\in ~\hat{\mathcal{B}}(\hat{r})}
||{L}- {L}^{*}||^2_{Fro},$$ conditioning on the event $$\min_{{L} \in \hat{\mathcal{B}}(\hat{r}),{S} \in \hat{\mathcal{A}}(\hat{s}),{\Sigma}={L}+{S}}{||{\Sigma}-{\Sigma}_n||^2_{Fro}}$$
under prescribed assumptions (see Theorem \ref{mine}). In fact we can write $$\min_{{L} ~\in ~\hat{\mathcal{B}}(\hat{r})}
||{L}- {L}^{*}||^2_{Fro}\leq \min_{{L} ~\in ~\hat{\mathcal{B}}(\hat{r})}||{L}- {Y}_{pre}||^2_{Fro}+ || {Y}_{pre}-{L^*}||^2_{Fro},$$
because ${Y}_{pre}$ is uniquely determined by the conditioning event. The same inequality holds in spectral norm.

Since it holds
$$||\hat{{L}}_{ALCE}- {L}^{*}||_2\leq ||\hat{{L}}_{UNALCE}-\hat{{L}}_{ALCE}||_{2}+||\hat{{L}}_{UNALCE}- {L}^{*}||_2,$$
we can write \begin{equation}0<||\hat{{L}}_{ALCE}- {L}^{*}||_2-||\hat{{L}}_{UNALCE}- {L}^{*}||_2\leq\breve{\psi}\end{equation}
given the conditioning event.
As a consequence, since $||\hat{{L}}_{UNALCE}~-~\hat{{L}}_{ALCE}||_{Fro} = tr(\hat{{L}}_{UNALCE}~-~\hat{{L}}_{ALCE})^2 =r\breve{\psi}^2$,
we obtain
\begin{equation}0<||\hat{{L}}_{ALCE}- {L}^{*}||_{Fro}-||\hat{{L}}_{UNALCE}- {L}^{*}||_{Fro}\leq\sqrt{r}\breve{\psi}.\end{equation}

The analogous triangular inequality for the sparse component is
$$||\hat{{S}}_{ALCE}- {S}^{*}||^2_{Fro}\leq ||\hat{{S}}_{UNALCE}-\hat{{S}}_{ALCE}||^2_{Fro}+||\hat{{S}}_{UNALCE}- {S}^{*}||^2_{Fro}.$$
In order to quantify $||\hat{{S}}_{UNALCE}-\hat{{S}}_{ALCE}||^2_{Fro}$, we need to study the behaviour of the term 
$\sum_{i=1}^{p} (\hat{l}_{UNALCE,ii}~-~\hat{l}_{ALCE,ii})^2$, which is less or equal to $r \breve{\psi}^2$, because it is less or equal to $tr(\hat{{L}}_{UNALCE}~-~\hat{{L}}_{ALCE})^2$. 

As a consequence, we have $||\hat{{S}}_{UNALCE}-\hat{{S}}_{ALCE}||_{Fro} \leq \sqrt{r} \breve{\psi}$.
Analogously to $\hat{{L}}_{UNALCE}$, we can prove that $$\hat{{S}}_{UNALCE}=\min_{{S}\in
\hat{\mathcal{A}}(\hat{s})}||{S}~-~ {S}^{*}||^2_{Fro},$$
conditioning on the event $$\min_{{L} \in \hat{\mathcal{B}}(\hat{r}),{S} \in \hat{\mathcal{A}}(\hat{s}),{\Sigma}={L}+{S}}{||{\Sigma}-{\Sigma}_n||^2_{Fro}}$$
under prescribed assumptions (see Theorem \ref{mine}).
In fact we can write $$\min_{{S} ~\in ~\hat{\mathcal{A}}_{diag}(\hat{s})}
||{S}- {S}^{*}||^2_{Fro}\leq \min_{{S} ~\in ~\hat{\mathcal{A}}_{diag}(\hat{s})}||{S}- {Z}_{pre}||^2_{Fro}+ || {Z}_{pre}-{S^*}||^2_{Fro},$$
because ${Z}_{pre}$ is uniquely determined by the conditioning event.

Therefore, we can write
\begin{equation}0<||\hat{{S}}_{ALCE}- {S}^{*}||_{Fro}-||\hat{{S}}_{UNALCE}- {S}^{*}||_{Fro}\leq \sqrt{r} \breve{\psi}.\end{equation}

The claim on $||\hat{{S}}_{UNALCE}- {S}^{*}||_{2}$ is less immediate.
We recall that $||\hat{{L}}_{UNALCE}-\hat{{L}}_{ALCE}||_{2} =||\hat{{U}}\breve{\psi}  {I}_r\hat{{U}'}||_{2}=\breve{\psi}$.
$\hat{{U}}\breve{\psi}  {I}_r\hat{{U}'}$ can be divided in the contribution coming from diagonal elements and the rest:
$||diag(\hat{{L}}_{UNALCE}-\hat{{L}}_{ALCE})+off-diag(\hat{{L}}_{UNALCE}-\hat{{L}}_{ALCE})||_{2}$.
Both contributes are part of $\hat{{U}}\breve{\psi}  {I}_r\hat{{U}'}$.

Given the matrix of eigenvectors $\hat{{U}}$, 
we can write $diag(\hat{{L}}_{UNALCE}-\hat{{L}}_{ALCE})=\sum_{i=1}^{p}||\hat{ u}'_{i}||^2{K}_{ii}$,
where ${K}_{ii}$ is a null matrix except for the $i$-th diagonal element equal to $\breve{\psi}$
and $\hat{ u}'_{i}$ is the $i$-th row of $\hat{{U}}$.
Similarly we can write $off-diag(\hat{{L}}_{UNALCE}-\hat{{L}}_{ALCE})=\sum_{i=1}^p\sum_{j\ne i} \hat{ u}'_{i} \hat{ u}_{j} {K}_{ij}$
where ${K}_{ij}$ is a null matrix except for the element $ij$ equal to $\breve{\psi}$.
Note that the rows of $\hat{{U}}$, differently from the columns, are not orthogonal.

Since all summands are orthogonal to each other (${A}\bot  {B} \Leftrightarrow tr({AB'})=0$), the triangular inequalities relative to $||diag(\hat{{L}}_{UNALCE}-\hat{{L}}_{ALCE})||$, $||off-diag(\hat{{L}}_{UNALCE}-\hat{{L}}_{ALCE})||$ and $||\hat{{L}}_{UNALCE}-\hat{{L}}_{ALCE}||_{2}$
become equalities.
Therefore we can write:
\begin{eqnarray}
||diag(\hat{{L}}_{UNALCE}-\hat{{L}}_{ALCE})||=\sum_{i=1}^{p}||\hat{ u}'_{i}||^2 \times ||{K}_{ii}||=\sum_{i=1}^{p}||\hat{ u}'_{i}||^2\breve{\psi}\\
||off-diag(\hat{{L}}_{UNALCE}-\hat{{L}}_{ALCE})||= \sum_{i=1}^p\sum_{j\ne i} \hat{ u}'_{i} \hat{ u}_{j}||{K}_{ij}||=\sum_{i=1}^p\sum_{j\ne i} \hat{ u}'_{i} \hat{ u}_{j}\breve{\psi}\\
||\hat{{L}}_{UNALCE}-\hat{{L}}_{ALCE}||_{2}=\sum_{i=1}^{p}||\hat{ u}'_{i}||^2 \times||{K}_{ii}||+\sum_{i=1}^p\sum_{j\ne i} \hat{ u}'_{i} \hat{ u}_{j}||{K}_{ij}||=\breve{\psi}.
\end{eqnarray}

From this consideration it follows that $$||diag(\hat{{L}}_{UNALCE}-\hat{{L}}_{ALCE})||\leq ||\hat{{L}}_{UNALCE}-\hat{{L}}_{ALCE}||_{2}=\breve{\psi}.$$
Since, by definition, $||diag(\hat{{S}}_{UNALCE}-\hat{{S}}_{ALCE})||=||diag(\hat{{L}}_{UNALCE}-\hat{{L}}_{ALCE})||$ (because $diag(\hat{{S}}_{UNALCE}-\hat{{S}}_{ALCE})=-diag(\hat{{L}}_{UNALCE}-\hat{{L}}_{ALCE})$), and recalling that $\hat{{S}}_{UNALCE}$ has the best approximation property (for Theorem \ref{mine}) given the conditioning event, we can conclude
\begin{equation}0<||\hat{{S}}_{ALCE}- {S}^{*}||_{2}-||\hat{{S}}_{UNALCE}- {S}^{*}||_{2}\leq \breve{\psi}.\end{equation}


\subsection{Proof of Corollary \ref{mine3}}
The relevant triangular inequality for the overall estimate is
$$||{\Sigma}_n-\hat{ {\Sigma}}_{ALCE}||_2\leq |||\hat{ {\Sigma}}_{UNALCE}-\hat{ {\Sigma}}_{ALCE}||_{2}+||{\Sigma}_n-\hat{{\Sigma}}_{UNALCE}||_2.$$
We know that, by definition, $||\hat{ {\Sigma}}_{UNALCE}-\hat{ {\Sigma}}_{ALCE}||_{2}=||off-diag(\hat{{L}}_{UNALCE}-\hat{{L}}_{ALCE})||_2$.
For the same considerations explained before, $$||off-diag(\hat{{L}}_{UNALCE}-\hat{{L}}_{ALCE})||\leq ||\hat{{L}}_{UNALCE}-\hat{{\Sigma}}_{ALCE}||_{2}=\breve{\psi}.$$
As a consequence, recalling that $\hat{{\Sigma}}_{ALCE}=\min_{{\Sigma}={L}+{S},{L} \in \hat{\mathcal{B}}(\hat{r}),{S}\in \hat{\mathcal{A}}(\hat{s})} ||{\Sigma}-{\Sigma}_n||^2_{Fro}$ under the described assumptions, we can conclude
\begin{equation}0<|| {\Sigma}_n-\hat{ {\Sigma}}_{ALCE}||_{2}-|| {\Sigma}_n-\hat{ {\Sigma}}_{UNALCE}||_{2}\leq \breve{\psi}.\label{2}\end{equation}
Since 
$||\hat{{L}}_{UNALCE}~-~\hat{{L}}_{ALCE}||^2_{Fro} = tr(\hat{{L}}_{UNALCE}~-~\hat{{L}}_{ALCE})^2=r\breve{\psi}^2$, we have
\begin{equation}0<|| off-diag(\hat{{L}}_{UNALCE}-\hat{{L}}_{ALCE})||_{Fro} \leq \sqrt{r}\breve{\psi}.\label{2}\end{equation}
We can then claim
\begin{equation}0<|| {\Sigma}_n-\hat{ {\Sigma}}_{ALCE}||_{Fro}-|| {\Sigma}_n-\hat{ {\Sigma}}_{UNALCE}||_{Fro}\leq \sqrt{r}\breve{\psi}.\label{fro}\end{equation}


Therefore, the real gain is terms of approximation of $ {\Sigma}_n$ respect to ALCE
measured in squared Frobenius norm is strictly positive and bounded from $r\breve{\psi}^2$.

\subsection{Proof of Theorem \ref{mine4}}

Conditioning on ${\Sigma}_n$, we can easily write $$||\hat{ {\Sigma}}_{UNALCE}- {\Sigma}^*||=$$
\begin{equation}=||\hat{ {\Sigma}}_{UNALCE}- {\Sigma}_n+ {\Sigma}_n- {\Sigma}^*|| \leq
||\hat{ {\Sigma}}_{UNALCE}- {\Sigma}_n|| + || {\Sigma}_n- {\Sigma}^*||. \label{pass}\end{equation}
The quality of the estimation input $|| {\Sigma}_n- {\Sigma}^*||$ does not depend on the estimation method. 


Therefore, by (\ref{2}) and (\ref{pass}),
it is straightforward that
\begin{equation}0 < ||\hat{ {\Sigma}}_{ALCE}- {\Sigma}^*||_{2}-||\hat{ {\Sigma}}_{UNALCE}- {\Sigma}^*||_{2} \leq \breve{\psi}.\label{SI2}
\end{equation}

Analogously, it is easy to prove that
\begin{equation}0 < ||\hat{ {\Sigma}}_{ALCE}- {\Sigma}^*||_{Fro}-||\hat{ {\Sigma}}_{UNALCE}- {\Sigma}^*||_{Fro} \leq \sqrt{r}\breve{\psi}.\label{SI}
\end{equation}

\subsection{Proof of Corollary \ref{mine5}}

We recall the following expression:
$$||(\hat{{L}}+\hat{{S}})^{*-1}-( {\Sigma})^{*-1}||_{Fro}= ||(\hat{{L}}+\hat{{S}})^{*-1}[\hat{{L}}+\hat{{S}}- {\Sigma}^*]( {\Sigma})^{*-1}|| \leq$$
$$\leq||(\hat{{L}}+\hat{{S}})^{*-1}||\cdot ||[\hat{{L}}+\hat{{S}}- {\Sigma}^*]||_{Fro}\cdot ||( {\Sigma})^{*-1}||.$$

From (\ref{SI})
we can conclude that
\begin{equation}0<||(\hat{{L}}_{ALCE}+\hat{{S}}_{ALCE})^{*-1}- {\Sigma}^{*-1}||_{Fro}-||(\hat{{L}}_{UNALCE}+\hat{{S}}_{UNALCE})^{*-1}- {\Sigma}^{*-1}||_{Fro} \leq  \sqrt{r}\breve{\psi}.\end{equation}

Analogously, since it holds
$$||(\hat{{L}}+\hat{{S}})^{*-1}-( {\Sigma})^{*-1}||= ||(\hat{{L}}+\hat{{S}})^{*-1}[\hat{{L}}+\hat{{S}}- {\Sigma}^*]( {\Sigma})^{*-1}|| \leq$$
$$\leq||(\hat{{L}}+\hat{{S}})^{*-1}||\cdot ||[\hat{{L}}+\hat{{S}}- {\Sigma}^*]||\cdot ||( {\Sigma})^{*-1}||.$$
it is straightforward that
\begin{equation}0<||(\hat{{L}}_{ALCE}+\hat{{S}}_{ALCE})^{*-1}- {\Sigma}^{*-1}||_{2}-||(\hat{{L}}_{UNALCE}+\hat{{S}}_{UNALCE})^{*-1}- {\Sigma}^{*-1}||_{2} \leq  \breve{\psi}.\end{equation}



\subsection{Proof of Corollary \ref{defin}}

We prove in sequence the three claims of the Corollary.
\begin{enumerate}
\item We start noting that $\hat{{L}}_{UNALCE}$, $\hat{{L}}_{ALCE}$ and ${U}_{ALCE}\breve{\psi} {I}_r {U}_{ALCE}'$ are $r$- ranked. We denote the respective spectral decompositions by: \begin{enumerate} \item $\hat{{B}}_{UNALCE}\hat{{B}}_{UNALCE}'$ with     $\hat{{B}}_{UNALCE}=\hat{{U}}_{ALCE}\sqrt{\hat{{D}}_{UNALCE}}$ ; \item $\hat{{B}}_{ALCE}\hat{{B}}_{ALCE}'$ with $\hat{{B}}_{ALCE}=\hat{{U}}_{ALCE}\sqrt{\hat{{D}}_{ALCE}}$; \item $({U}_{ALCE}\sqrt{\breve{\psi}})({U}_{ALCE}\sqrt{\breve{\psi}})'.$\end{enumerate}

As a consequence, by Lidskii dual inequality we note that
\begin{eqnarray}
{\lambda}_r(\hat{{L}}_{UNALCE})={\lambda}_r(\hat{{L}}_{ALCE}+\hat{{U}}_{ALCE}\breve{\psi} {I}_r \hat{{U}}_{ALCE}')=\nonumber\\
{\lambda}_r(\hat{{U}}_{ALCE}\hat{{D}}_{ALCE} \hat{{U}}_{ALCE}+\hat{{U}}_{ALCE}\breve{\psi} {I}_r \hat{{U}}_{ALCE}')={\lambda}_r(\hat{{L}}_{ALCE})+\breve{\psi}, \nonumber
\end{eqnarray}
which proves the claim on $\hat{{L}}_{UNALCE}$.
\item By Lidskii dual inequality, we note that
\begin{eqnarray}
{\lambda}_p(\hat{{S}}_{UNALCE})={\lambda}_p(\hat{{S}}_{ALCE}-diag(\hat{{U}}_{ALCE}\breve{\psi} {I}_r \hat{{U}}_{ALCE}'))\geq\nonumber\\
\geq{\lambda}_p(\hat{{S}}_{ALCE})+{\lambda}_p(-diag(\hat{{U}}_{ALCE}\breve{\psi} {I}_r \hat{{U}}_{ALCE}')).\nonumber
\end{eqnarray}
The matrix $-diag(\hat{{U}}_{ALCE}\breve{\psi} {I}_r \hat{{U}}_{ALCE}')$ is a $p$-dimensional squared matrix having as $i-$th element the quantity $-||{u}'_i||^2\breve{\psi}$,
where ${u}'_i$, $i=1,\ldots,p$, is the i-th row of the matrix $\hat{{U}}_{ALCE}$.
Since $tr(-diag(\hat{{U}}_{ALCE}\breve{\psi} {I}_r \hat{{U}}_{ALCE}'))=tr(-\hat{{U}}_{ALCE}\breve{\psi} {I}_r \hat{{U}}_{ALCE}')=-r\breve{\psi}$,
it descends that\\ ${\lambda}_p(-diag(\hat{{U}}_{ALCE}\breve{\psi} {I}_r \hat{{U}}_{ALCE}'))\leq \frac{r}{p} \breve{\psi}$, i.e. 
\begin{equation}-\frac{r}{p}\breve{\psi} \leq {\lambda}_p(-diag(\hat{{U}}_{ALCE}\breve{\psi} {I}_r \hat{{U}}_{ALCE}'))\leq 0.\nonumber\end{equation} 
Therefore we obtain
\begin{equation}
{\lambda}_p(\hat{{S}}_{UNALCE})\geq{\lambda}_p(\hat{{S}}_{ALCE})-\frac{r}{p}\breve{\psi},\nonumber
\end{equation}
which proves the claim on $\hat{{S}}_{UNALCE}$.
\item By Lidskii dual inequality, we note that
\begin{eqnarray}
{\lambda}_p(\hat{{\Sigma}}_{UNALCE})={\lambda}_p(\hat{{\Sigma}}_{ALCE}+\hat{{U}}_{ALCE}\breve{\psi} {I}_r \hat{{U}}_{ALCE}'-diag(\hat{{U}}_{ALCE}\breve{\psi} {I}_r \hat{{U}}_{ALCE}'))\geq\nonumber\\
\geq{\lambda}_p(\hat{{\Sigma}}_{ALCE})+{\lambda}_p(\hat{{U}}_{ALCE}\breve{\psi} {I}_r \hat{{U}}_{ALCE}')-{\lambda}_p(diag(\hat{{U}}_{ALCE}\breve{\psi} {I}_r \hat{{U}}_{ALCE}')).\nonumber
\end{eqnarray}
Recalling the argument above and noting that\\ ${\lambda}_p(\hat{{U}}_{ALCE}\breve{\psi} {I}_r \hat{{U}}_{ALCE}')=0$ because $rank(\hat{{U}}_{ALCE}\breve{\psi} {I}_r \hat{{U}}_{ALCE}')=\hat{r}$, it descends
\begin{equation}
{\lambda}_p(\hat{{\Sigma}}_{UNALCE})\geq{\lambda}_p(\hat{{\Sigma}}_{ALCE})+0-\breve{\psi}={\lambda}_p(\hat{{\Sigma}}_{ALCE})-\frac{r}{p}\breve{\psi},\nonumber
\end{equation}
which proves the claim on $\hat{{\Sigma}}_{UNALCE}$.
\end{enumerate}

\bibliographystyle{chicago}
\bibliography{unshr2top}

\begin{thebibliography}{}

\bibitem[\protect\citeauthoryear{Agarwal, Negahban, and Wainwright}{Agarwal
  et~al.}{2012}]{agarwal2012noisy}
Agarwal, A., S.~Negahban, and M.~J. Wainwright (2012).
\newblock Noisy matrix decomposition via convex relaxation: Optimal rates in
  high dimensions.
\newblock {\em The Annals of Statistics\/}, 1171--1197.

\bibitem[\protect\citeauthoryear{Anderson}{Anderson}{1984}]{anderson1984multivariate}
Anderson, T. (1984).
\newblock Multivariate statistical analysis.
\newblock {\em Wiley and Sons, New York, NY\/}.

\bibitem[\protect\citeauthoryear{Bai}{Bai}{2003}]{bai2003inferential}
Bai, J. (2003).
\newblock Inferential theory for factor models of large dimensions.
\newblock {\em Econometrica\/}~{\em 71\/}(1), 135--171.

\bibitem[\protect\citeauthoryear{Bai and Ng}{Bai and
  Ng}{2002}]{bai2002determining}
Bai, J. and S.~Ng (2002).
\newblock Determining the number of factors in approximate factor models.
\newblock {\em Econometrica\/}~{\em 70\/}(1), 191--221.

\bibitem[\protect\citeauthoryear{Bickel and Levina}{Bickel and
  Levina}{2008a}]{bickel2008covariance}
Bickel, P.~J. and E.~Levina (2008a).
\newblock Covariance regularization by thresholding.
\newblock {\em The Annals of Statistics\/}, 2577--2604.

\bibitem[\protect\citeauthoryear{Bickel and Levina}{Bickel and
  Levina}{2008b}]{bickel2008regularized}
Bickel, P.~J. and E.~Levina (2008b).
\newblock Regularized estimation of large covariance matrices.
\newblock {\em The Annals of Statistics\/}, 199--227.

\bibitem[\protect\citeauthoryear{Cai, Cand{\`e}s, and Shen}{Cai
  et~al.}{2010}]{cai2010singular}
Cai, J.-F., E.~J. Cand{\`e}s, and Z.~Shen (2010).
\newblock A singular value thresholding algorithm for matrix completion.
\newblock {\em SIAM Journal on Optimization\/}~{\em 20\/}(4), 1956--1982.

\bibitem[\protect\citeauthoryear{Cai and Liu}{Cai and
  Liu}{2011}]{cai2011adaptive}
Cai, T. and W.~Liu (2011).
\newblock Adaptive thresholding for sparse covariance matrix estimation.
\newblock {\em Journal of the American Statistical Association\/}~{\em
  106\/}(494), 672--684.

\bibitem[\protect\citeauthoryear{Cai, Zhang, and Zhou}{Cai
  et~al.}{2010}]{cai2010}
Cai, T.~T., C.-H. Zhang, and H.~H. Zhou (2010, 08).
\newblock Optimal rates of convergence for covariance matrix estimation.
\newblock {\em The Annals of Statistics\/}~{\em 38\/}(4), 2118--2144.

\bibitem[\protect\citeauthoryear{Chandrasekaran, Parrilo, and
  Willsky}{Chandrasekaran et~al.}{2012}]{chandrasekaran2012}
Chandrasekaran, V., P.~A. Parrilo, and A.~S. Willsky (2012, 08).
\newblock Latent variable graphical model selection via convex optimization.
\newblock {\em The Annals of Statistics\/}~{\em 40\/}(4), 1935--1967.

\bibitem[\protect\citeauthoryear{Chandrasekaran, Sanghavi, Parrilo, and
  Willsky}{Chandrasekaran et~al.}{2011}]{chandrasekaran2011rank}
Chandrasekaran, V., S.~Sanghavi, P.~A. Parrilo, and A.~S. Willsky (2011).
\newblock Rank-sparsity incoherence for matrix decomposition.
\newblock {\em SIAM Journal on Optimization\/}~{\em 21\/}(2), 572--596.

\bibitem[\protect\citeauthoryear{Clarke}{Clarke}{1990}]{clarke1990optimization}
Clarke, F.~H. (1990).
\newblock {\em Optimization and nonsmooth analysis}.
\newblock SIAM.

\bibitem[\protect\citeauthoryear{Daubechies, Defrise, and De~Mol}{Daubechies
  et~al.}{2004}]{daubechies2004iterative}
Daubechies, I., M.~Defrise, and C.~De~Mol (2004).
\newblock An iterative thresholding algorithm for linear inverse problems with
  a sparsity constraint.
\newblock {\em Communications on pure and applied mathematics\/}~{\em
  57\/}(11), 1413--1457.

\bibitem[\protect\citeauthoryear{Davidson and Szarek}{Davidson and
  Szarek}{2001}]{davidson2001local}
Davidson, K.~R. and S.~J. Szarek (2001).
\newblock Local operator theory, random matrices and banach spaces.
\newblock {\em Handbook of the geometry of Banach spaces\/}~{\em 1\/}(317-366),
  131.

\bibitem[\protect\citeauthoryear{Dey and Srinivasan}{Dey and
  Srinivasan}{1985}]{dey1985estimation}
Dey, D.~K. and C.~Srinivasan (1985).
\newblock Estimation of a covariance matrix under stein's loss.
\newblock {\em The Annals of Statistics\/}, 1581--1591.

\bibitem[\protect\citeauthoryear{Fan, Fan, and Lv}{Fan
  et~al.}{2008}]{fan2008high}
Fan, J., Y.~Fan, and J.~Lv (2008).
\newblock High dimensional covariance matrix estimation using a factor model.
\newblock {\em Journal of Econometrics\/}~{\em 147\/}(1), 186--197.

\bibitem[\protect\citeauthoryear{Fan, Liao, and Liu}{Fan
  et~al.}{2016}]{fan2016overview}
Fan, J., Y.~Liao, and H.~Liu (2016).
\newblock An overview of the estimation of large covariance and precision
  matrices.
\newblock {\em The Econometrics Journal\/}~{\em 19\/}(1).

\bibitem[\protect\citeauthoryear{Fan, Liao, and Mincheva}{Fan
  et~al.}{2011}]{fan2011high}
Fan, J., Y.~Liao, and M.~Mincheva (2011).
\newblock High dimensional covariance matrix estimation in approximate factor
  models.
\newblock {\em The Annals of Statistics\/}~{\em 39\/}(6), 3320--3356.

\bibitem[\protect\citeauthoryear{Fan, Liao, and Mincheva}{Fan
  et~al.}{2013}]{fan2013large}
Fan, J., Y.~Liao, and M.~Mincheva (2013).
\newblock Large covariance estimation by thresholding principal orthogonal
  complements.
\newblock {\em Journal of the Royal Statistical Society: Series B (Statistical
  Methodology)\/}~{\em 75\/}(4), 603--680.

\bibitem[\protect\citeauthoryear{Farn\'{e}}{Farn\'{e}}{2016}]{farne2016large}
Farn\'{e}, M. (2016).
\newblock {\em Large Covariance Matrix Estimation by Composite Minimization}.
\newblock Ph.\ D. thesis, Alma Mater Studiorum.

\bibitem[\protect\citeauthoryear{Farn\'{e} and Montanari}{Farn\'{e} and
  Montanari}{2018}]{dataset}
Farn\'{e}, M. and A.~Montanari (2018).
\newblock A large covariance matrix estimator under intermediate spikiness
  regimes.
\newblock \url{https://data.mendeley.com/datasets/nh97vfvhkt}.

\bibitem[\protect\citeauthoryear{Fazel}{Fazel}{2002}]{fazel2002matrix}
Fazel, M. (2002).
\newblock {\em Matrix rank minimization with applications}.
\newblock Ph.\ D. thesis, PhD thesis, Stanford University.

\bibitem[\protect\citeauthoryear{Fazel, Hindi, and Boyd}{Fazel
  et~al.}{2001}]{fazel2001rank}
Fazel, M., H.~Hindi, and S.~P. Boyd (2001).
\newblock A rank minimization heuristic with application to minimum order
  system approximation.
\newblock In {\em American Control Conference, 2001. Proceedings of the 2001},
  Volume~6, pp.\  4734--4739. IEEE.

\bibitem[\protect\citeauthoryear{Friedman, Hastie, and Tibshirani}{Friedman
  et~al.}{2008}]{friedman2008sparse}
Friedman, J., T.~Hastie, and R.~Tibshirani (2008).
\newblock Sparse inverse covariance estimation with the graphical lasso.
\newblock {\em Biostatistics\/}~{\em 9\/}(3), 432--441.

\bibitem[\protect\citeauthoryear{Furrer and Bengtsson}{Furrer and
  Bengtsson}{2007}]{furrer2007estimation}
Furrer, R. and T.~Bengtsson (2007).
\newblock Estimation of high-dimensional prior and posterior covariance
  matrices in kalman filter variants.
\newblock {\em Journal of Multivariate Analysis\/}~{\em 98\/}(2), 227--255.

\bibitem[\protect\citeauthoryear{Lam et~al.}{Lam
  et~al.}{2016}]{lam2016nonparametric}
Lam, C. et~al. (2016).
\newblock Nonparametric eigenvalue-regularized precision or covariance matrix
  estimator.
\newblock {\em The Annals of Statistics\/}~{\em 44\/}(3), 928--953.

\bibitem[\protect\citeauthoryear{Ledoit and Wolf}{Ledoit and
  Wolf}{2004}]{ledoit2004well}
Ledoit, O. and M.~Wolf (2004).
\newblock A well-conditioned estimator for large-dimensional covariance
  matrices.
\newblock {\em Journal of multivariate analysis\/}~{\em 88\/}(2), 365--411.

\bibitem[\protect\citeauthoryear{Ledoit and Wolf}{Ledoit and
  Wolf}{2015}]{ledoit2015spectrum}
Ledoit, O. and M.~Wolf (2015).
\newblock Spectrum estimation: A unified framework for covariance matrix
  estimation and pca in large dimensions.
\newblock {\em Journal of Multivariate Analysis\/}~{\em 139}, 360--384.

\bibitem[\protect\citeauthoryear{Luo}{Luo}{2011a}]{luo2011high}
Luo, X. (2011a).
\newblock High dimensional low rank and sparse covariance matrix estimation via
  convex minimization.
\newblock {\em Arxiv preprint\/}.

\bibitem[\protect\citeauthoryear{Luo}{Luo}{2011b}]{luo2011recovering}
Luo, X. (2011b).
\newblock Recovering model structures from large low rank and sparse covariance
  matrix estimation.
\newblock {\em arXiv preprint arXiv:1111.1133\/}.

\bibitem[\protect\citeauthoryear{Nesterov}{Nesterov}{2013}]{nesterov2013gradient}
Nesterov, Y. (2013).
\newblock Gradient methods for minimizing composite functions.
\newblock {\em Mathematical Programming\/}~{\em 140\/}(1), 125--161.

\bibitem[\protect\citeauthoryear{Qiu and Chen}{Qiu and
  Chen}{2015}]{qiu2015bandwidth}
Qiu, Y. and S.~X. Chen (2015).
\newblock Bandwidth selection for high-dimensional covariance matrix
  estimation.
\newblock {\em Journal of the American Statistical Association\/}~{\em
  110\/}(511), 1160--1174.

\bibitem[\protect\citeauthoryear{Rockafellar}{Rockafellar}{2015}]{rockafellar2015convex}
Rockafellar, R.~T. (2015).
\newblock {\em Convex analysis}.
\newblock Princeton university press.

\bibitem[\protect\citeauthoryear{Rothman, Levina, and Zhu}{Rothman
  et~al.}{2009}]{rothman2009generalized}
Rothman, A.~J., E.~Levina, and J.~Zhu (2009).
\newblock Generalized thresholding of large covariance matrices.
\newblock {\em Journal of the American Statistical Association\/}~{\em
  104\/}(485), 177--186.

\end{thebibliography}

\end{document}